\documentclass[pre,floatfix,superscriptaddress]{revtex4}
\usepackage{graphicx}
\usepackage{epic,eepic}
\usepackage{graphics}
\usepackage{epsfig}
\usepackage{subfigure}

\renewcommand{\r}{{\rho}}

\newcommand{\T}{T}

\newcommand{\half}{\frac{1}{2}}

\newcommand{\dt}{{\Delta t}}
 
\newcommand{\pii}{{\partial_i}}

\newcommand{\pj}{{\partial_j}}

\newcommand{\pt}{{\partial_t}}

\newcommand{\be}{\begin{equation}}
\newcommand{\ee}{\end{equation}}
\newcommand{\bea}{\begin{eqnarray}}
\newcommand{\eea}{\end{eqnarray}}

\newcommand{\RB}{Rayleigh-B\'enard}
\newcommand{\RT}{Rayleigh-Taylor}

\usepackage{amsmath}
\usepackage{amsfonts}
\usepackage{amssymb}
\usepackage{bm}
\usepackage{color}
\usepackage{graphicx}

\begin{document}
\title{ Lattice Boltzmann Methods for thermal flows: continuum  limit  and applications to compressible \RT\ systems.}

\author{A. Scagliarini} \affiliation{Department of Physics and INFN,
  University of Tor Vergata,\\Via della Ricerca Scientifica 1, 00133
  Rome, Italy\\and International Collaboration for
  Turbulence Research} \author{L. Biferale} 
\affiliation{Department of Physics and INFN,
  University of Tor Vergata,\\Via della Ricerca Scientifica 1, 00133
  Rome, Italy\\and International Collaboration for
  Turbulence Research} \author{M. Sbragaglia}
\affiliation{Department of Physics and INFN, University of Tor
  Vergata, \\ Via della Ricerca Scientifica 1, 00133 Rome, Italy}
\author{K. Sugiyama} \affiliation{Department of Mechanical Engineering, School of Engineering,\\ The University of Tokyo, 7-3-1, Hongo, Bunkyo-ku, Tokyo 113-8656, Japan} \author{F. Toschi}
\affiliation{Department of Physics and Department of Mathematics and
  Computer Science, Eindhoven University of Technology, 5600 MB
  Eindhoven, The Netherlands; and International Collaboration for
  Turbulence Research}

\begin{abstract}
  We compute the continuum thermo-hydrodynamical limit of a new
  formulation of lattice kinetic equations for thermal compressible
  flows, recently proposed in  [Sbragaglia et al., {\it J. Fluid
    Mech.} {\bf 628} 299 (2009)].  We show that the hydrodynamical
  manifold is given by the correct compressible  Fourier-Navier-Stokes
  equations for a perfect fluid. We validate the numerical algorithm
 by means of exact results for transition to convection
  in \RB\  compressible systems and against direct
  comparison with finite-difference schemes. The method is stable and
 reliable up to temperature jumps between top and
  bottom walls
 of the order of $50\%$ the averaged bulk temperature.  We use this method to study \RT\ instability for
  compressible stratified flows and we determine the growth of the
  mixing layer at changing Atwood numbers up to $At \sim 0.4$. 
We highlight the role played by the adiabatic gradient in stopping the
mixing layer growth in presence of high stratification and we
quantify  the asymmetric growth rate for spikes and bubbles for two
dimensional \RT\ systems with resolution up to $L_x \times
L_z = 1664 \times 4400$ and with Rayleigh numbers up to $ Ra \sim 2
\times 10^{10}$. 

\end{abstract}

\maketitle
\section{Introduction}
Lattice implementations of discrete-velocity kinetic models have
gained considerable interest in the last decades, as efficient tools
for the theoretical and computational investigation of the physics of
complex flows
\cite{SC1,Yeo,Wagb,HeLuo,Ladd,Hartingb,ladd-review,jos}. 
 An important class
of discrete-velocity models for ideal fluid flows, the lattice
Boltzmann models (LBM) \cite{Gladrow,BSV,Chen} can be derived from the
continuum Boltzmann (BGK) equation \cite{BGK54}, upon expansion in
Hermite velocity space of the single particle distribution function,
$f({\bm x}, {\bm \xi},t)$, describing the probability of finding a
molecule at space-time location $({\bm x},t)$ and with velocity ${\bm
  \xi}$ \cite{HeLuo,ShanHe98,Martys98,Shan06}.  As a result, the
corresponding lattice dynamics acquires a systematic justification in
terms of an underlying continuum kinetic theory. The state-of-the-art
is satisfactory concerning iso-thermal flows, even in presence of
complex bulk physics (multi-phase, multi-components)
\cite{SC1,Yeo,HeDoolen01} and/or with complex boundary conditions
 such as roughness, non-wetting walls and slip-length 
\cite{Hartingb,pre.nostro,prl.nostro,Harting}.\\
The situation is much less satisfactory when temperature plays an
active role in the flow evolution, due to complex compressible effects
which are present even in ideal fluid/gas or to phase-change in
multi-phase systems, or both. Only a few years ago, one could frankly
admit that not a single known Lattice Boltzmann approach could handle,
in a realistic way, thermal problems properly. The main
difficulties being the development of subtle instabilities when the
local velocity increases. In the last years, the situation has started
to improve, with different attempts being made to describe active
thermal modes within a fully discretized Boltzmann approach
\cite{LL,Karlin,Prasianakis,Sofonea,Gonnella,watari,Shan06,Philippi06,Nie08,Meng09}. These
studies show that in order to recover the right
continuum descriptions with the correct symmetries for the
internal energy flux, one needs to enlarge the number of
discrete speeds (a possible choice, for space filling schemes following a Gauss-Hermite quadrature \cite{Shan06,Philippi06}, is $37$ speeds in 2d \cite{Shan07,Philippi06} and $107$ speeds in 3d \cite{Surmas09}), or to add ad-hoc counter-terms canceling spurious anisotropic operators \cite{Karlin,Prasianakis}. Otherwise, different hybrid attempts have been proposed, where temperature evolution is solved using finite difference methods  \cite{LL} or with lattice schemes able to reproduce thermal Van der Waals fluids in the continuum limit \cite{Gonnella}. Boundary conditions  \cite{Sofonea,boundary} and stability issues \cite{Siebert08} are also much more involved when thermal modes are present.  It is fair to say that not a single model emerged as the optimal choice, and only a few explorative studies have been performed in order to check potentiality
and limitations of each proposed solution.

The aim of this paper is twofold.  First, we intend to further discuss
a recent formulation, proposed by some of us in \cite{JFM.nostro}, for
a new way to incorporate the effects of external/internal forces in
thermal LBM. We provide here the full explicit Chapman-Enskog
expansion, whose results where only anticipated without proof in
\cite{JFM.nostro}, in order to show the convergence of the model to
the Fourier-Navier-Stokes equations.  We validate the method in a case
where thermal compressible effects play a major role, i.e. the
transition to convection in a compressible \RB\ system of height
$L_z$, with an imposed temperature jump, $T_u-T_d= \Delta T$. 
For such systems, it is possible to calculate the critical Rayleigh
number analytically \cite{spiegel1} at changing both the
stratification parameter (also known as the scale height), $Z=\Delta
T/T_u$, and the polytropic index, $m= g/(R \beta) -1$, where $R$ is
the gas constant, $g$ the gravity acceleration and $\beta=\Delta
T/L_z$  the temperature gradient. We show here that our LBM scheme
is able to handle temperature jumps as high as $\Delta T/T_u =2$ for
both positive and negative values of the polytropic index (stable and
unstable density stratification). Such systems are clearly very far
from the classical Oberbeck-Boussinesq approximation
\cite{spiegel3,Lohse}.

Second, we study highly compressible \RT\ systems, for the initial
configuration where two blobs of the same fluids are prepared with two
different temperatures (hot, less dense, blob below, cold, denser,
blob above). We show that the method is able to handle the highly
non-trivial spatio-temporal evolution of the system even in the
developing turbulent phase. In this case, we could push the numerics
up to Atwood numbers $At \sim 0.4$. Maximum Rayleigh numbers achieved
are $Ra \sim 4
\times 10^{10}$ for $At=0.05$ and $Ra \sim 2
\times 10^{9}$ for $At=0.4$. We present results on: (i) the
growth of the mixing-layer at changing the compressibility degree,
including the asymmetry in the quadratic growth of spikes and bubbles
dynamics; (ii) a new effect of stratification which stops the mixing
length growth when a critical width, $L_{ad}$ is reached. We interpret
this as due to the existence of the {\it adiabatic gradient}: when the
jumps between the two moving fronts leads to a temperature gradient,
$\Delta T/L_{ad}$, of the order of the {\it adiabatic gradient} the
dynamics stops and only thermal diffusive mixing may further acts.

Technically speaking, the main novelty of the thermal-LBM formulation
proposed in \cite{JFM.nostro} relies on the fact that it is possible
to incorporate the effects of an external and/or internal force
(gravity and/or intermolecular potential) via a suitable shift of both
momentum and temperature appearing in the local equilibrium
distribution of the Boltzmann collision operator. Doing that, the
systems acquires an elegant self-consistent formulation and a stable
spatio-temporal evolution also in presence of compressible effects, as
demonstrated by the examples anticipated before and detailed later.

The paper is organized as follows. In Section \ref{sec:kinetic} we
briefly remind the details of the LBM formulation and we discuss the
first result of this paper: the continuum thermo-hydrodynamical limit,
given by the Fourier-Navier-Stokes eqs., as obtained from a rigorous
Chapman-Enskog expansion of the discrete model.
In Section \ref{sec:transition} we show first the validation of the
discretized algorithm by studying the transition to convection in
compressible \RB\ systems and comparing the results with exact
analytical calculations, at changing the scale height and the
polytropic index.  In the same section we also present validation of
the method against finite-difference methods, for the same set-up
but after the transition, once convective rolls are present and stationary.  In
Section \ref{sec:RT} we show the investigations of another non-trivial
compressible case: \RT\ system, for two different Atwood numbers
$At=0.05$ and $At=0.4$.  Conclusions and perspectives close the paper in
Section \ref{sec:conclusions}.
  
\section{Thermal Kinetic Model and Continuum Theory}
\label{sec:kinetic}
The main goal of this section is to show how to use a Thermal-LBM to
discretize continuum thermal kinetic equations in presence of
internal/external forces and how to extract via a suitable
Chapman-Enskog multiscale expansion the relative hydrodynamical
evolution, given in term of the {\it forced} Fourier-Navier-Stokes
equations. The first issue was already discussed in \cite{JFM.nostro}:
here we  briefly recall it and then discuss the second issue in
details.

A Thermal-Kinetic description of a compressible gas/fluid of variable
density, $\rho$, local velocity ${\bm u }$, internal energy, ${\cal
  K}$ and subjected to a local body force density, ${\bm g}$, is given in the
continuum by the following set of equations (repeated indices are
summed upon):
\begin{equation}\label{eq:1}
\begin{cases}
  \partial_t \rho + \partial_i ( \rho u_i) = 0  \\
  \partial_t (\rho u_k) + \partial_i (P_{ik}) = \rho g_k  \\
  \partial_t {\cal K} + \frac{1}{2} \partial_i q_i = \rho g_i u_i
\end{cases}
\end{equation}
where $P_{ik}$ and $q_i$ are momentum and energy fluxes (still unknown
at this level of description).

In \cite{JFM.nostro}, it is shown that it is possible to recover
exactly the above set of equations, starting from a continuum
Boltzmann Equations and introducing a suitable shift of the velocity
and temperature fields entering in the local equilibrium:
$f^{(eq)}({\bm \xi}; \rho, T, {\bm u}) \rightarrow { f}^{(eq)}({\bm
  \xi}; \rho, {\bar T}, {\bm {\bar u}})$.  The new --shifted--
Boltzmann formulation being:
\begin{eqnarray} \label{MASTERSHIFT3}
&& \frac{\partial f }{\partial t}+{\bm \xi} \cdot {\bm \nabla} f=-\frac{1}{\tau}(f- {\ f}^{(eq)}); \\
&& { f}^{(eq)}({\bm \xi}; \rho, \bar T, \bar {\bm u})=\frac{\rho}{(2 \pi {\bar T})^{D/2}} e^{-|{\bm \xi}- {\bm {\bar u}}|^2/2  {\bar  T}}.
\end{eqnarray}
Where the shifted local velocity and
temperature must take the following form:
\begin{equation}
\label{eq:shift2}
\bar {\bm u} = {\bm u} + \tau {\bm g} \qquad \bar T = \T -  \tau^2 g^2/D.
\end{equation}
The
lattice counterpart of the continuum description (\ref{MASTERSHIFT3})
can be obtained through the usual lattice Boltzmann discretization:
$$ f_{l}({\bm x}+ {\bm c}_{l} \dt,t+\dt) - f_{l}({\bm x},t) =
-\frac{\dt}{\tau}(f_{l}({\bm x},t) - f_l^{(eq)})  
$$
where the equilibrium 
is expressed in terms of  hydrodynamical fields on the lattice, 
${f}_{l}^{(eq)}({\bm  x},\rho^{(L)},\bar{\bm  u}^{(L)},\bar\T^{(L)})$, and   the subscript $l$ runs over the discrete
 set of velocities,
${\bm c}_l$. The superscript $L$ indicates that the macroscopic
fields are now defined in terms of the lattice Boltzmann populations:
\be
\begin{cases}
  \rho^{(L)}  = \sum_l f_l;\\
  \rho^{(L)}  {\bm u}^{(L)}   = \sum_l {\bm c}_l f_l; \\
  D \rho^{(L)} \T^{(L)} = \sum_l \left|{\bm c}_l - {\bm
      u}^{(L)}\right|^2 f_l.
\end{cases}
\ee 
In \cite{JFM.nostro} it was shown that the lattice version of the
shifted fields entering in the Boltzmann equilibrium (see Appendix A
for its detailed form) is:
$$
\bar{\bm u}^{(L)}= {\bm u}^{(L)} + \tau {\bm g} \qquad \bar{\T}^{(L)}=
\T^{(L)} + \frac{\tau(\dt-\tau) g^2}{D } + {\cal O} (\Delta t)^2.
$$

As it is known, lattice discretizations induce non trivial corrections
terms in the macroscopic evolution of averaged hydrodynamical
quantities. In particular both momentum and temperature must be
renormalized by discretization effects in order to recover the correct
thermal kinetic description (\ref{eq:1}) out of the discretized LBM
variables.  Density is left unchanged, $\rho^{(H)} = \rho$, while the
first non trivial correction to momentum is given by the pre and
post-collisional average \cite{Buick00,Guo02}: 
\be
\label{eq:shift_lattice} {\bm u}^{(H)}={\bm u}^{(L)}+\frac{\Delta
  t}{2} {\bm g} \ee and the first non-trivial, correction to the
temperature field by \cite{JFM.nostro}: \be {\T}^{(H)} = \T^{(L)} +
\frac{(\dt)^2g^2}{4 D }.
\label{eq:thydro}
\ee
 
Using this {\it renormalized} hydrodynamical fields, one recover by a
suitable Taylor expansions in $\Delta t$ the thermo-hydrodynamical
equations \cite{JFM.nostro}:
\begin{equation}
\begin{cases}
  \pt \r^{(H)} + \partial_i ( \r u_i^{(H)}) = 0\\
  \pt (\r^{(H)} u_k^{(H)}) + \partial_i (P_{ik}^{(H)}) = \rho^{(H)} g_k\\
  \pt {\cal K}^{(H)} + \half \partial_i q_i^{(H)} = \rho^{(H)} g_i
  u_i^{(H)}.
\end{cases}
\label{eq:hydro}
\end{equation}

The above equations are still unclosed. A closure
ansatz to express the stress tensor, $P_{ik}^{(H)}$, and the heat flux,
$q_i^{(H)}$, in terms of lower order moments is needed. This ends our short
review of the backup material.\\
We proceed now with a systematic multi-scale closure of
(\ref{eq:hydro}) in order to control the small wave-length limit where
the full Fourier-Navier-Stokes equations emerge. The main added value
with respect to previous similar calculations \cite{Siebert07} is the explicit inclusion of the effects of the external force ${\bm g}$ in the Chapman-Enskog expansion. \\
In order to perform the calculations, we need to introduce a hierarchy
of temporal and spatial scales, via the introduction of a small
parameter, $\epsilon$:
$$
\partial_{t} \rightarrow
\epsilon \partial_{1t}+\epsilon^2 \partial_{2t} ; \qquad
\partial_{i} \rightarrow \epsilon \partial_{i}
$$
and the corresponding expansion for the Boltzmann distributions
$$
f=f^{(0)}+\epsilon f^{(1)}+\epsilon^2 f^{(2)}+\epsilon^3
f^{(3)}+\epsilon^4 f^{(4)}+.....
$$
together with a suitable rescaling of the forcing terms, ${g} \sim {\cal O}(\epsilon)$ \cite{Buick00}. The various rescalings immediately reflect in the explicit expansion of the equilibrium distribution in terms of Hermite polynomials, ${\cal H}^{(n)}_{l}$:
$$
f^{(eq)}_{l}=w_l \sum_{n} \frac{1}{n!} {\bm a}^{(n)}_0 {\cal H}^{(n)}_{l}
$$
where $w_l$ are suitable weights \cite{Shan07,Nie08}. The projections on the different Hermite polynomials, ${\bm a}^{(n)}_0$, are explicitly given in Appendix A.\\
After a long calculation, fully detailed in the Appendix, one shows
that the leading long wavelength limit coincides with the continuum
Fourier-Navier-Stokes equations of an ideal compressible gas given by:
\be
\label{eq:FNS1}
\begin{cases}
  \partial_{t} \rho+ \partial_i (\rho u_i^{(H)})=0\\
\rho \partial_{t}  u_j^{(H)}+\rho u_i^{(H)} \partial_i u_j^{(H)}=\rho g_j+
\partial_i   \sigma_{ij}^{(H)}\\
\rho \partial_{t} e^{(H)}+\rho u_i^{(H)} \partial_i e^{(H)}=
\sigma_{ij}^{(H)} \partial_j u_i^{(H)}+\partial_i (k \partial_i
e^{(H)}) ;
\end{cases}
\ee
with the ideal gas internal energy given by: $ e^{(H)}=\frac{D}{2}
\T^{(H)}$. The stress tensor is given by:
$$
  \sigma_{ij}^{(H)}=-\rho \T^{(H)}  \delta_{ij}+\nu ( \partial_i  u_j^{(H)}+
  \partial_j u_i^{(H)})+ \delta_{ij} \left(\xi-\frac{\nu}{c_v}
  \right) \partial_k u_k^{(H)}.
$$
The shear and bulk  viscosities are:
$$
\nu=\T^{(H)} \rho \left( \tau-\frac{\Delta t}{2} \right);\hspace{.2in}
\left( \xi-\frac{\nu}{c_v} \right)=-\frac{\T^{(H)} \rho}{c_v} \left(
  \tau-\frac{\Delta t}{2} \right)
$$
and the thermal conductivity: \be k=c_p \T^{(H)} \rho \left(
  \tau-\frac{\Delta t}{2} \right).  \ee These are therefore the
equations for a compressible gas with an ideal equation of state: \be
p =\rho \T^{(H)} \ee and ideal specific heats: \be c_v=\frac{D}{2};
\hspace{.2in} c_p=\frac{D}{2}+1 \ee It is not difficult to show that
in the case the external forces are conservative, written in a
potential form depending only on the density, one may easily
incorporate these effects in the definition of an internal energy,
opening the way to discuss also non-ideal equations of state
\cite{JFM.nostro}.

\section{Transition to convection in \RB\ compressible systems}
\label{sec:transition}
A first non trivial application of the above algorithm can be found
studying the behavior of \RB\ cells both considering the effects of
compressibility and stratification to the transition from diffusive to
convective dynamics \cite{spiegel1,gauthier,robinson} or to the case
of fully turbulent non-Oberbeck-Boussinesq convection
\cite{Lohse}. Here we concentrate on the first issue (see top panel of
figure \ref{fig:rt} for a schematic view), results on high Rayleigh
turbulent convection will be published elsewhere.  First, let us
rewrite the set of equations (\ref{eq:FNS1}) in a more transparent
way, dropping for simplicity the superscript $H$ in all variables and
using the explicit expression of the internal energy in term of the
temperature field: \be
\begin{cases}
  D_t \rho = - \rho \pii u_i \\
  \rho D_t u_i = - \pii p - \rho g\delta_{i,z} + \eta \partial_{jj} u_i + \left(1-\frac{1}{c_v}\right) \eta \pii \pj u_j \\
  \rho c_v D_t T + p \pii u_i = k \partial_{ii} T + \eta (\partial_i u_j+\partial_j u_i -\frac{1}{c_v} \delta_{ij} \partial_k u_k)\partial_i u_j
\end{cases}
\ee where we have introduced the material derivative, $D_t
= \partial_t + u_j\pj$, and we have assumed constant viscous and
thermal conductivity coefficients \cite{spiegel1,note}.  The equation
of state is, $p =\rho T$, i.e. it is given in terms of quantities
normalized such that the gas constant is $R=1$. For a cell of height
$L_z$ and with imposed bottom and top temperature, $T_{d}$ and
$T_{u}$, the hydrostatic equilibrium is easily found in terms of the
temperature jump across the cell, $\beta = (T_{d}-T_{u})/L_z = \Delta
T/L_z$: \be
\label{eq:hydrostatic}
\begin{cases}
  T_0(z) = (T_{d}+T_u)/2 - \beta\, z\\
  \rho_0(z) = \tilde \rho\, (T_0(z)/\tilde T)^m\\
  p_0(z) = \tilde p\,  (T_0(z)/\tilde T)^{m+1}\\
\end{cases}
\ee 
where the two integration constants must satisfy, $\tilde p =
\tilde \rho \tilde T$, with $\tilde T$ a reference temperature,
$\tilde T = (T_u+T_d)/2$. In (\ref{eq:hydrostatic}) we have introduced
also the polytropic index: $m = g/\beta - 1$. At changing the
polytropic index, one changes the hydrostatic profiles of density and
pressure. In order to be unstable, the profile must obviously verify,
$\beta >0$ (if $g>0$, as assumed here) and therefore the interesting
polytropic interval is limited to $ m \ge - 1$. Furthermore, unstable
fluctuations may develop only if the hydrostatic temperature gradient,
$\beta$ is larger than the {\it adiabatic gradient}, $\beta_{ad} =
g/c_p$, i.e. only when the adiabatic transformation of a hot/cold spot
of fluid moving up/down induces a temperature variation that does not
exceed the hydrostatic change \cite{landau}. This limits the
interesting interval excursion of the polytropic index from above, $m
< c_p-1$, which in our units, for an ideal gas in 2d, means $m
<1$. The limitation from above is a typical important example induced
by compressibility/stratification, i.e. by the fact that a cold/hot
fluid spots may contracts or expand during their spatio-temporal
evolution. Stratification can be also measured by the {\it scale
  height}, i.e. a typical length scale, $L_h$, built in terms of mean
hydrostatic quantities. In our case, the most natural way to define it
is by using the temperature profile: $L_h = (T_u/\Delta T) L_z =
L_z/Z$. Where we  used the dimensionless parameter, $Z =
\Delta T/T_u$ which is a direct measurement of the stratification
effects: for $Z \gg 1$, the cell height $L_z$ is much larger than the
typical stratification length, i.e. the fluid is highly stratified. On
the other hand, the limit $Z \rightarrow 0$ corresponds to the
so-called Oberbeck-Boussinesq approximation, where both stratification
and compressibility 
are vanishingly small. The latter is, by far, the
most studied convection configuration, even though some important
applications for astrophysics \cite{conv-astro,conv-astro1} 
and recently also for
laboratory set-up \cite{conv-lab1,conv-lab2,conv-lab3} cannot neglect
compressible modes.  It is possible to show \cite{spiegel3} that in
the Boussinesq approximation, the dependency from the polytropic index
disappear (as it must obviously do) while it remains a possible effect
induced by the adiabatic gradient (usually small on laboratory
experiments, but not necessarily on atmospheric scales).\\

\begin{figure}
  \advance\leftskip-0.55cm
  \includegraphics[width=0.8\textwidth]{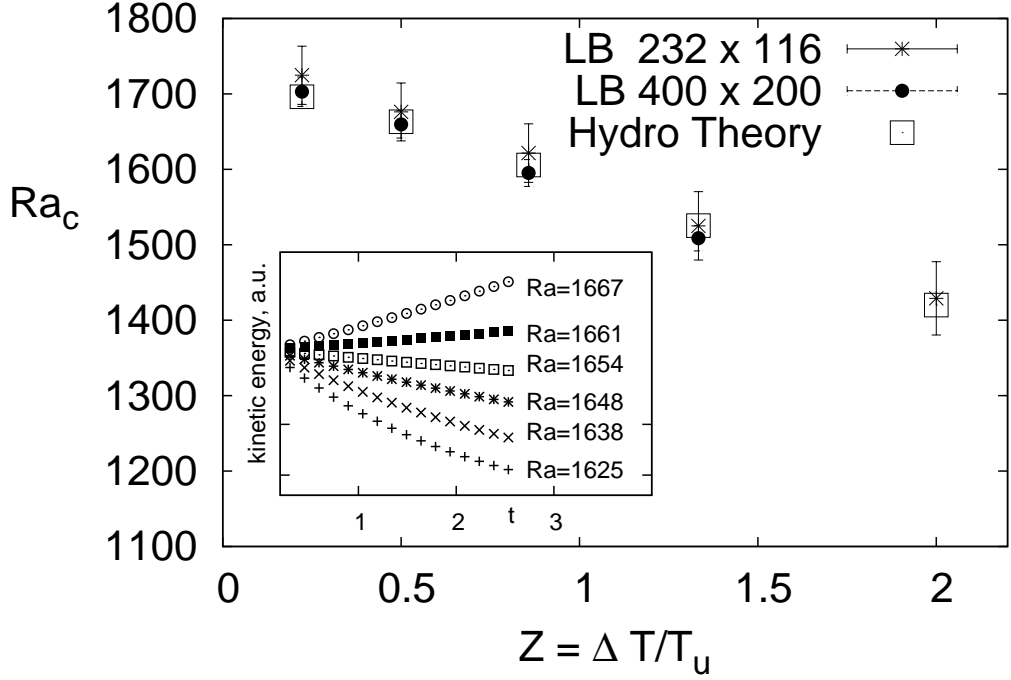}
  \caption{Critical Rayleigh number, estimated at the center of the
    cell, ${\tilde Ra_c}$, at changing the polytropic index, $m$, the
    scale height, $Z$, and the numerical resolutions. For the smallest
    resolution, $L_x\times L_z = 232 \times 116$, the plotted values
    corresponds to (a) $Z=0.22, \Delta T = 0.2, T_d= 0.9, m = -0.942$; (b) $Z=0.5, \Delta T = 0.4, T_d= 0.8, m = -0.971$; (c) $Z=0.86, \Delta T = 0.6, T_d =
    0.7, m = -0.9806$; (d) $Z=1.33, \Delta T = 0.8, T_d = 0.6, m = -0.9855$; 
    (e) $Z=2.0, \Delta T = 1.2, T_d = 0.6, m =-0.990$. Theoretical
    values are obtained solving the linearized equations as described
    in \cite{spiegel1}.  Inset: time evolution of the total kinetic
    energy (in arbitrary units) for Rayleigh numbers lower and higher than the critical one for the parameter case (c). The unit of time corresponds to $10000$ LB integration steps.}
  \label{fig:Rac}
\end{figure}

\begin{figure}
\begin{center}
  \advance\leftskip-0.75cm
  \includegraphics[width=0.6\textwidth]{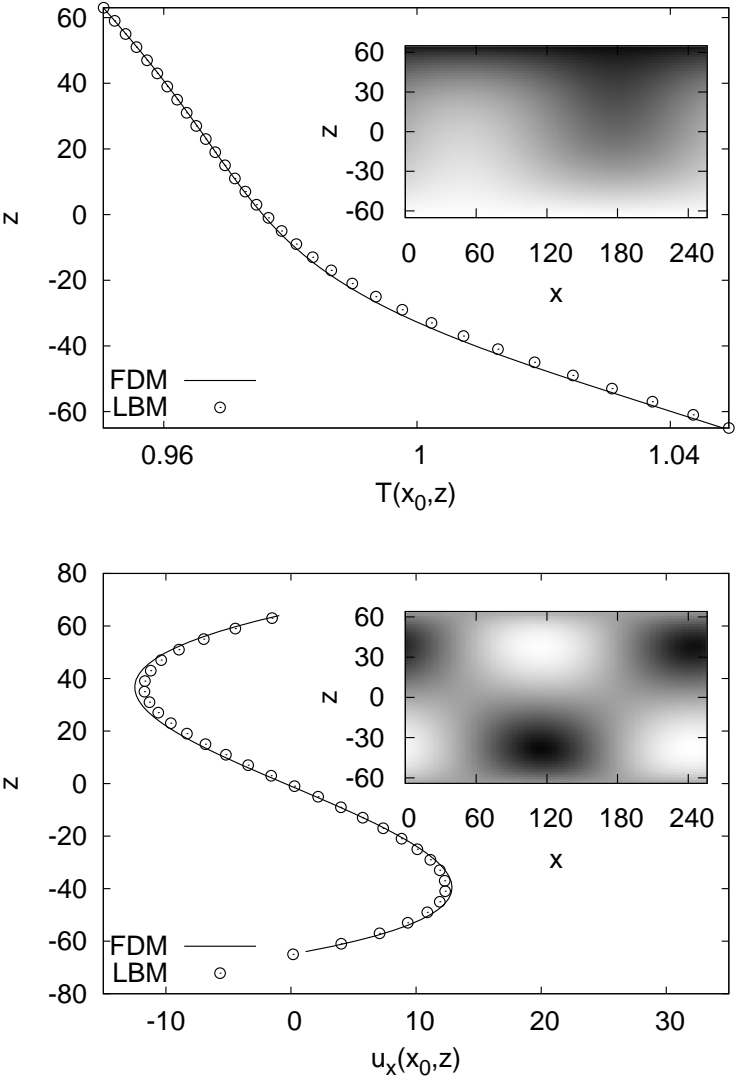}
  \caption{Comparison between one-dimensional vertical cut of the
    stationary temperature and velocity profiles after transition to a
    convective two-rolls configuration. Up: $T(x_0,z)$ at $x_0$ such
    that $x_0=0.69 L_x$. Circles  correspond to the Lattice Boltzmann
    algorithm (LBM); solid line corresponds  to a finite difference
    calculations (FDM)  \cite{kazu1,kazu2}. Down:
    the same of above plot but for the stream-wise velocity,
    $u_x(x_0,z)$. In the insets we show a grey-scale coded representations of
    the convective stationary rolls in the whole two-dimensional 
domain (up: temperature; down: stream-wise
    velocity). }
  \label{fig:Mauro-Kazu}
\end{center}
\end{figure}

We use this complex set-up to benchmark the thermal-LBM algorithm
proposed, and probe its robustness at changing compressibility.  This
can be done directly against exact results on the emergence of
convective instability in the system. It is possible to
calculate, in a closed form, the stability problem of the linearized
system around the hydrostatic solution (\ref{eq:hydrostatic}),
 for both slip or no-slip
velocity boundary conditions and for any polytropic index
\cite{spiegel1}: these are just suitable extensions of the well known Rayleigh calculation made for the incompressible case \cite{chandra}. \\
Stratification makes the problem non-homogeneous (in the vertical
direction) and therefore it is not possible to define in a unique way
the Rayleigh number. Anyhow, it turns out that it is possible to
introduce a height-dependent Rayleigh number which rules the 
linearized problem: 
\be
\label{eq:Ra}
Ra(z) =
\frac{(g/T_0(z))L_z^4(\beta-\beta_{ad})}{(k/\rho_0(z)c_p)(\nu/\rho_0(z))},
\ee and one can express the whole bifurcation diagram in terms the
value of the Rayleigh number at a given height, say the middle of the
cell $z=L_z/2$ for example: $ \tilde Ra = Ra(L_z/2) $. Different works
have been devoted to the calculations of the critical $\tilde Ra_c$ at
changing the polytropic index, the scale height, $Z$ and the boundary
conditions at the top/bottom plates \cite{spiegel1,gough,graham}.  A
result of the stability calculation predicts that there exists a
critical Rayleigh number which depends only on the polytropic index, $m$,
on the stratification parameter, $Z$, and on the wavelength, $a$, of
the perturbation, $\tilde Ra_c(m,Z,a)$.  The hydrostatic solution will
therefore become unstable under perturbation of a wavelength
corresponding to the minimum possible critical Rayleigh
number. Compressibility and stratification may have different
effects, either stabilizing or destabilizing the systems, depending on
the hydrostatic underlying equilibrium. For example, if the
hydrostatic profile has an unstable density profile, $m<0$, one gets that the critical Rayleigh decreases at increasing
temperature jumps. The opposite happens when density is stably
stratified, $m>0$.
From the definition of Rayleigh given in (\ref{eq:Ra}), it is easy to
realize the importance of the adiabatic gradient, $\beta_{ad} = g/c_p$,
i.e. if $\beta <\beta_{ad}$, the control parameter is always negative
and the system will always be linearly stable.
In figure \ref{fig:Rac} we show the result of a numerical search of the
critical Rayleigh number (i.e. the onset of the transition to
convection) using our LBM algorithm, obtained by exploring the long
time behavior of the system, prepared with a small perturbation to its
hydrostatic equilibrium, and monitoring the successive temporal
growth/decline of the total kinetic energy (example in the inset).
The LBM has been applied by imposing no-slip
impenetrable boundary conditions for the velocity field at top/bottom
walls, $u_z(z=\pm L_z/2)=0$; $u_x({z=\pm L_z/2})=0$; and
with an imposed constant temperature jump, $T({z=- L_z/2}) = T_d$;
$T({z= L_z/2}) = T_u$. Lateral boundaries are fully
periodic. Technical details on the way to implement the given boundary
conditions in the LBM algorithm are given in Appendix B.  In the same
figure we also report the critical Rayleigh numbers obtained from the
LBM exploration, compared with the exact analytical results obtained
by solving numerically the eigenvalue problem for the linearized equations as
given in \cite{spiegel1}. As one can see, the agreement is good, even
for large temperature jumps, up to $ Z \sim 2 $. Larger values of $Z$
are difficult to reach, because of limitations imposed by numerical
stability of the boundary conditions and by the growth of
unstable compressible modes in the system. In order to overcome such
limitation one should probably extend the Hermite projections to
higher and higher orders \cite{Siebert08}. 
 The main error source in
the determination of the critical Rayleigh number out of our LBM
method stems from the presence of spurious, small, departure from the
exact linear profile in the mean temperature close to the boundary
walls. This departure goes together with the existence of small
spurious transverse velocity for two-three grid layers close to the
wall and are due to the existence of discrete velocities which
connects up to three layers in the lattice inducing non-local boundary
conditions effects (see appendix A and B for details). Such effects
can be annoying for the investigation of highly turbulent regimes,
where the boundary layer dynamics becomes crucial to drive the correct
thermal exchange with the bulk \cite{verzicco}.  This shortcoming can
be strongly reduced by moving from LBM algorithms using exact
streaming (as done here) to LBM based on finite-volume schemes \cite{CHEN}. Details in this direction will be published elsewhere. The small spurious oscillations close to the boundaries does not prevent to get a very good quantitative validation of the algorithms also when large scale convective rolls are present.  For example in figure \ref{fig:Mauro-Kazu} we make a one-to-one comparison
of the LBM numerics with a numerical study using finite-difference
scheme for incompressible \RB\ systems \cite{kazu1,kazu2}.  Again, the
stationary profiles are perfectly superposing, as shown for both
temperature and velocity in figure \ref{fig:Mauro-Kazu}. This ends our
validation section. In the next section we apply the new algorithm to
study  compressible dynamics, as it is the case
of \RT\ instabilities in thermal stratified flows. In the latter case,
the small spurious oscillations close to the walls are obviously
completely unimportant, being the bulk the only physically interesting
region.

\section{\RT\ systems}
\label{sec:RT}
\begin{figure}
\begin{center}
  \includegraphics[scale=0.95]{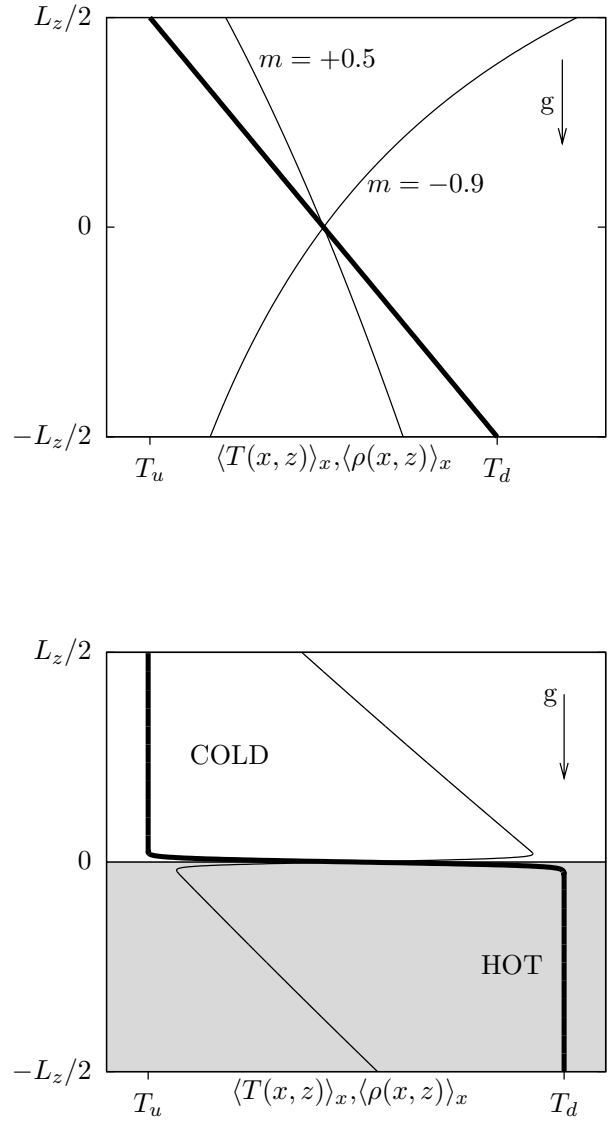}
  \caption{Upper panel: \RB\ geometry and set-up of the initial
    configuration given by eq. (\ref{eq:hydrostatic}); two cases with $m=+0.5$ 
    and $m=-0.9$.  On the horizontal axis we show the mean temperature and
    density profiles as a function of the $z$-height (plotted on the vertical axis). 
The bold and
    tiny solid lines represent the temperature and density profiles
    respectively. Lower panel: \RT\ initial configuration given by eq. 
(\ref{eq:hydroRT}). Bold and tiny lines as in the upper panel. }
\label{fig:rt}
\end{center}
\end{figure}

\begin{figure}
  \begin{center}
    \advance\leftskip-2.9cm
    \includegraphics[width=0.9\textwidth]{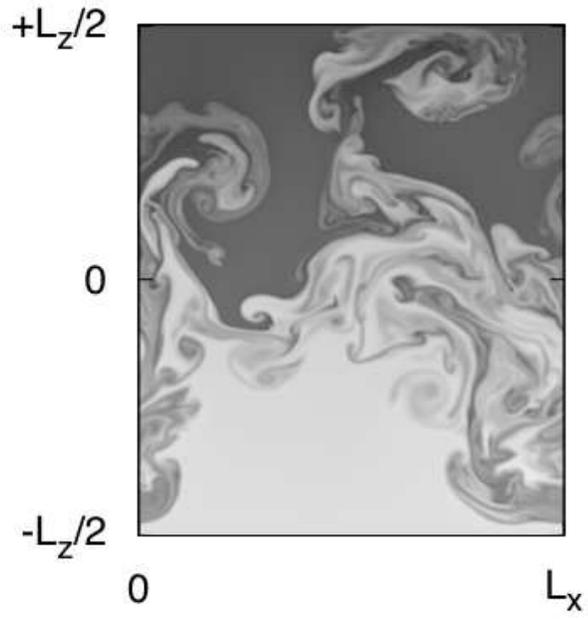}
    \caption{Spatial configuration  for a typical RT run with
      $L_x\times L_z = 800 \times 2400$, $T_u=0.95$, $T_d=1.05$ at
      time $t = 4 \tilde{\tau}$ (run A in table I).}
    \label{fig:rt.evolution}
  \end{center}
\end{figure}

Superposition of a heavy fluid above a lighter one in a constant
acceleration field depicts a hydro-dynamic unstable configuration
called the \RT\ (RT) instability \cite{chandra} with applications on
different fields going from inertial-confinement fusion \cite{rt2} to
supernovae explosions \cite{rt3} and many others \cite{rt.review}.
Although this instability was studied for decades it is still an open
problem in several aspects \cite{rt4}. In particular, it is crucial to
control the initial and late evolution of the mixing layer between the
two miscible fluids; the small-scale turbulent fluctuations, their
anisotropic/isotropic ratio; their dependency on the initial
perturbation spectrum or on the physical dimensions of the embedding
space \cite{jot,boffi}. In many cases, especially concerning
astrophysical and nuclear applications, the two fluids evolve with
strong compressible and/or stratification effects, a situation which
is difficult to investigate either theoretically or numerically. Here,
we concentrate on the large scale properties of the mixing layer,
studying a slightly different RT system than what usually found in the
literature: the spatio temporal evolution of a single component fluid
when initially prepared on the hydrostatic unstable equilibrium,
i.e. with a cold uniform region in the top half and a hot uniform
region on the bottom half (see bottom panel of figure \ref{fig:rt}).
For the sake of simplicity we limit the investigation to the 2d
case. While small-scales fluctuations may be strongly different in 2d
or 3d geometries, the large scale mixing layer growth is not supposed
to change its qualitative evolution \cite{chertkov,cabot}.
A  grey-scale coded snapshot of a typical RT run is shown in
figure \ref{fig:rt.evolution} showing all the complexity of the
phenomena.  Let us start to define precisely the initial set-up. We
prepare a single component compressible flow in a 2d tank of size,
$L_x \times L_z$, with adiabatic and no-slip boundary conditions on
the top and bottom walls, and with periodic boundary conditions on the
vertical boundaries. For convenience we define the initial interface
to be at height $z=0$, the box extending up to $z=L_z/2$ above and
$z=-L_z/2$ below it (see figure \ref{fig:rt}). In the two half volumes
we then fix two different homogeneous temperature, with the
corresponding hydrostatic density profiles, $\rho_0$, verifying
\cite{bernstein}: \be
\label{eq:phydroRT}
\partial_z p_0(z) = -g \rho_0(z).  \ee Considering that in each half
we have $p_0(z) =T \rho_0(z)$,with $T$ fixed, the solution has an
exponentially decaying behavior in the two half volumes, each one
driven by its own temperature value.  The initial hydrostatic unstable
configuration is therefore given by: 
\be
\label{eq:hydroRT}
\begin{cases}
  T_0(z) = T_{u};\,\, \rho_0(z) = \rho_u \exp(-g(z-z_c)/T_u);\qquad z >0\\
  T_0(z) = T_{d};\,\, \rho_0(z) = \rho_b \exp(-g(z-z_c)/T_d);\qquad z <0.\\
\end{cases}
\ee To be at equilibrium, we require to have the same pressure at the
interface, $z=z_c=0$; which translates in a simple condition on the
prefactor of the above expressions: \be \rho_u T_u = \rho_b T_d.  \ee
Because $T_u<T_d$, we have at the interface $\rho_u> \rho_b$.  As far
as we know, there are no exhaustive detailed calculations of the
stability problem for such configuration, even though not too
different from the usual RT compressible case
\cite{chandra,gauthier3,gauthier.priv}.  As said, this is not the
common way to study RT systems, which is usually meant as the
superposition of two different miscible fluids, isothermal, with
different densities \cite{chandra,gauthier2,gauthier3,jot}. As far as
compressible effects are small, one may safely neglect pressure
fluctuations and write -- for the case of an ideal gas: \be
\frac{\delta \rho}{\rho} \sim -\frac{\delta T}{T} \ee and the two RT
experiments are then strictly equivalent. Moreover, in the latter
case, if one may neglect the dependency of viscosity and thermal
diffusivity from temperature, the final evolution is indistinguishably
from the evolution of the temperature in the Boussinesq approximation
\cite{chertkov,boffi}. 
Here we will study both the case of small
compressibility and small stratification, where pressure is always
close to its hydrostatic value, $p \sim p_0 $, and the case when
compressibility becomes dynamically relevant, changing the global
large scale evolution of the mixing layer.
\subsection{RT instability in thermally active flows:
the role of the adiabatic gradient}
\begin{figure}
\begin{center}
  \advance\leftskip-0.55cm
  \includegraphics[width=0.7\textwidth]{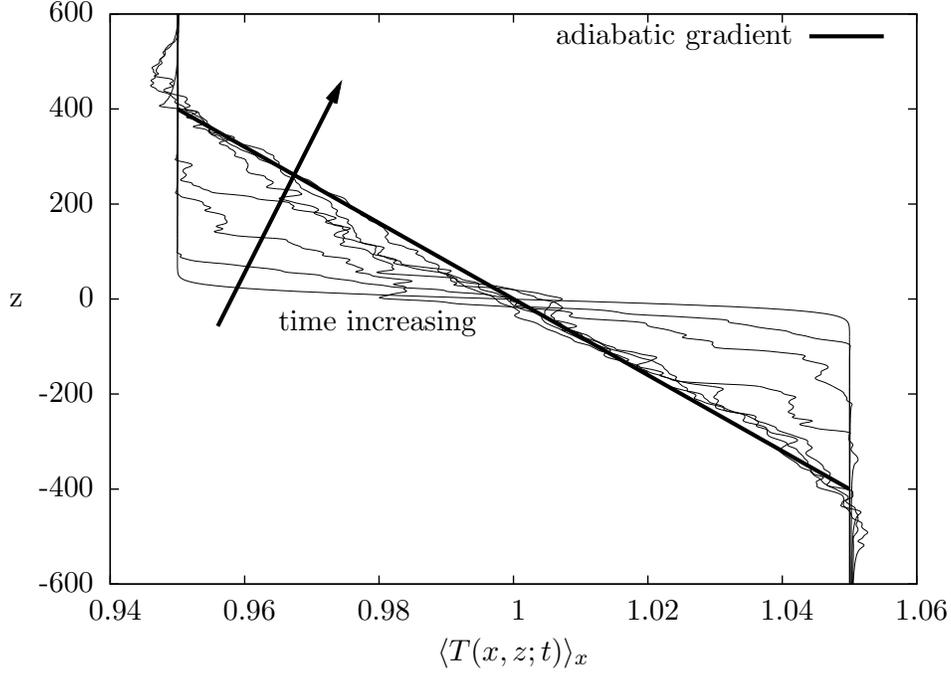}
  \caption{Temporal evolution of the mean temperature profile,
    $\langle T(x,z,t) \rangle_x$ at changing time, $t = n \delta t$,
    with $\delta t \approx 1.5 \tau_{ad} $ $n=1,2,\dots,7$. Notice that the
    profile approaches more and more the linear behaviour dictated by
    the adiabatic gradient, $ \langle T(x,z,t) \rangle_x = (T_u+T_d)/2
    - z g/c_p$. Time is adimensionalized by using a reference time
    based on adiabatic quantities, $\tau_{ad} = \sqrt{L_{ad}/(g \,
      \Delta T/T_u)}$ }
  \label{fig:adiabatic}
\end{center}
\end{figure}

The main novelty in the set up here investigated is due to the
presence of new effects induced by the adiabatic gradient,
which in our case can be written as in the previous Section
$\beta_{ad} = g/c_p$. In order to understand the main physical point
it is useful to think at the RT mixing layer as equivalent to a
(developing) Rayleigh-B\'enard system with an imposed mean temperature
gradient \cite{celani1,celani2}.  Let us denote with $L_{ml}(t)$ the typical
width of the RT mixing layer at a given time as measured for example
from the distance between the two elevations where the mean
temperature profile is $1\%$ lower or higher then the bottom and top,
respectively, unmixed temperature values, $L_{ml} = z_u-z_d$, where
$\langle T (x,z_{u}) \rangle_x = 1.01 T_u$ and $\langle T (x,z_{d})
\rangle_x = 0.99 T_d$.  It is well known that the temperature tends to
develop a linear profile inside the mixing region, the resulting
instantaneous temperature gradient is then given by $\beta(t) =
(T_d-T_u)/L_{ml}(t)$, and it decreases in time inversely to the growth
of the mixing length.  As a result, soon or later (if the box is tall
enough) the instantaneous temperature gradient will become of the same
order of the adiabatic gradient, $\beta(t) \sim \beta_{ad}$ and the
growth of the mixing length will stop. One can define an instantaneous
Rayleigh number, driving the physics inside the mixing layer,
estimated as in Section \ref{sec:transition}: \be
\label{eq:Rat}
\tilde Ra(t) = \frac{(g/\tilde T_0) L_{ml}^4(t)
  (\beta(t)-\beta_{ad})}{(k/ \tilde \rho_0 c_p)(\nu/ \tilde \rho_0)},
\ee where $\tilde{ (\cdot)}$ indicates quantities evaluated at the
middle layer. It is clear that for small times, $\beta(t) \ll
\beta_{ad}$, the effective instantaneous Rayleigh number is high: the
system is unstable, and the mixing length grows. On the other hand, as
time elapses, the vertical mean temperature gradient decreases, until
a point when, $\beta(t)\sim \beta_{ad}$, the instantaneous effective
Rayleigh number becomes $\tilde Ra(t) \sim {\cal O}(1)$ and the system
tends to be stabilized. We can then identify an {\it adiabatic
  length}: $$L_{ad} = (T_d-T_u)/\beta_{ad} = c_p \Delta T/g$$ which
determines the maximum length achievable by the mixing layer, in our
configuration. Let us notice that in absence of the adiabatic
gradient, the Rayleigh number would continue to grow indefinitely,
being proportional to the third power of  $L_{ml}(t)$, as it is the case for usual RT
systems.  If the profile coinciding with the adiabatic gradient is
going to be fully stable depends on the top/bottom boundary conditions
imposed on the whole spatial domain. In any case, when temperature
matches the adiabatic profile, the system strongly feel it, showing a
sudden slowing down of the mixing layer growth. To our knowledge, this
effect has never been predicted before, within this framework.
We show in figure \ref{fig:adiabatic} the evolution of temperature
profiles when adiabatic effects are important. It is clear how the
mixing layer growth is strongly slowed down when $L_{ml}(t) \sim
L_{ad}$; afterward only very slow relaxation process happens further,
mainly at the border between the edge of the mixing layer and the
fluids region with homogeneous temperature.\\
A possible way to estimate quantitatively when and how the adiabatic
gradient starts to play a role in the growth of the mixing length is
to use a simple phenomenological closure for large scale quantities in
the system.  We start from the self-similar scaling predicted by
\cite{cook,cook2} for the homogeneous not stratified growth: \be
\label{eq:cook}
(\dot L_{ml}(t))^2 = 4 \alpha^{(L)}\, g\, At\, L_{ml}(t) 
\ee
which has a unique solution (beside the trivial one, $L_{ml}=0$) in terms of the initial value, $L_{ml}(t_0)$: 
\be
\label{eq:Lt5}
L_{ml}(t) = L_{ml}(t_0) +  2\,\sqrt{L_{ml}(t_0) \, \alpha^{(L)}\, At\, g}\,  (t-t_0)  + \alpha^{(L)}\, At\,g\, (t-t_0)^2.
\ee
Eq. (\ref{eq:cook}) offers the advantage to be local in time, i.e. one may extract the value of $\alpha^{(L)}$ by a simple evaluation of the plateau in the ratio $(\dot L_{ml})^2/L_{ml}$, time by time. In order to minimally modify the above expression considering the saturation effects induced by stratification, we propose to use:
\be
\label{eq:luca-andrea}
(\dot L_{ml}(t))^2 = 4 \alpha^{(L)}\, g\, At\, L_{ml}(t)\psi\left(\frac{L_{ml}(t)}{L_{ad}}\right) 
\ee
where $\psi=\psi(x)$ must be a function fulfilling the condition $\psi \rightarrow 1$ as $ x \rightarrow 0$ 
(that is for $L_{ad} \rightarrow \infty$), in order to recover the equation (\ref{eq:cook}) for the not stratified case
when the adiabatic gradient goes to zero. We further add the  
requirement of reaching the adiabatic profile with
zero velocity and acceleration, enforcing
 a strict irreversible growth, i.e. $\dot L_{ml} \ge 0$, as it must
 be for the case of miscible fluids. Under these assumptions,  
 it can be shown that the simplest form for the function $\psi$ is:  
\be
\label{eq:luca-andrea1}
\psi\left(\frac{L}{L_{ad}}\right) = C\left[e^{-\left(\frac{L-L_{ad}}{L_{ad}}\right)}-\left(\frac{2L_{ad}-L}{L_{ad}}\right)\right]
\ee
where the prefactor $C$ must be set equal to $1/(e-2)$ to comply with the prescribed boundary conditions.
Equation (\ref{eq:luca-andrea}) must be considered as a zero-th order phenomenological way to take into account of the  adiabatic gradient in the mixing layer evolution.

\begin{figure}
\begin{center}
\advance\leftskip-0.55cm
\includegraphics[width=0.7\textwidth]{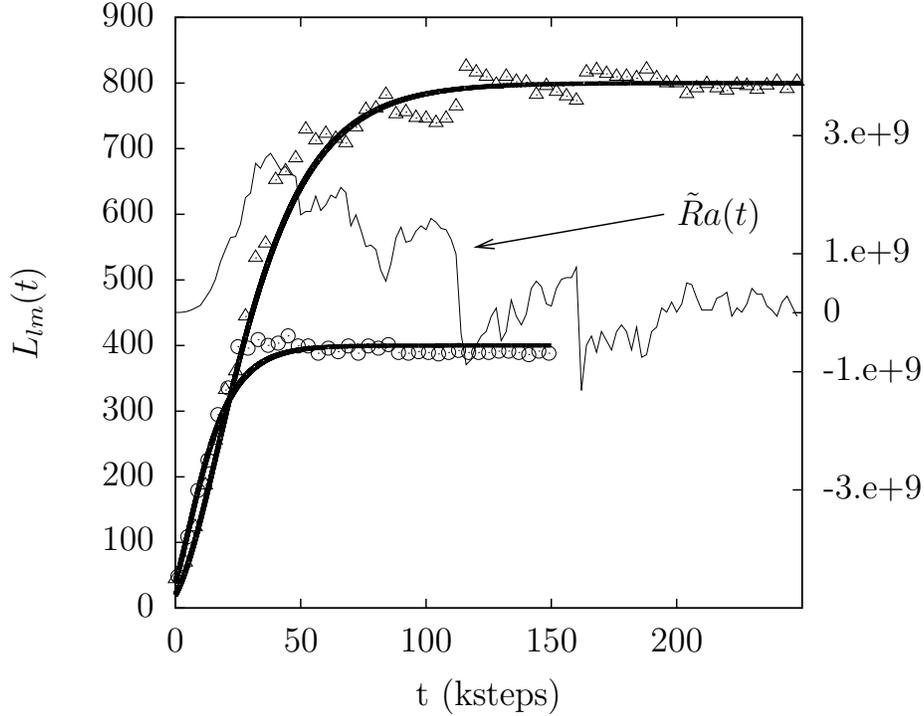}
\caption{Evolutions of the mixing layer, $L_{ml}(t)$ versus time with two different adiabatic lengths: (a): 
 $L_{ad}=800$, $g=2.5\; 10^{-4}$ (triangles); 
(b) $L_{ad}=400$, $g=5\; 10^{-4}$ (circles); 
Both cases have $At=0.05$, $\nu=0.001$ and $\kappa = 0.002$. Solid bold lines correspond to the 
theoretical prediction (\ref{eq:luca-andrea}) with $\alpha^{(L)}
=0.05$. Continuous line correspond to the evolution of the
instantaneous Rayleigh number (\ref{eq:Rat}) calculated for case (a), scale on the right y-axis.}

\label{fig:adiabatic.profile}
\end{center}
\end{figure}

We integrated numerically eq. (\ref{eq:luca-andrea}) testing the
result in figure \ref{fig:adiabatic.profile} where we show that 
it is possible to fit the global evolution of the mixing length
$L_{ml}(t)$, by using reasonable \cite{rt4} values of $\alpha^{(L)}$, for all times,
including the long time behavior where $L_{ml}(t) \sim L_{ad}$.  In
the same figure, we also show the behaviour of the time-dependent
effective Rayleigh number (\ref{eq:Rat}), estimated using the
instantaneous mixing length, $L_{ml}(t)$. As one can see, after the
initial monotonic growth of the turbulent intensity, there appear a
sudden slowing down, as identified by a strong reduction in the
effective Rayleigh number.  We can therefore safely assume that the
solution of our equation (\ref{eq:luca-andrea}) is a good
generalization of (\ref{eq:Lt5}) including also the adiabatic
gradients effects.
\subsection{Compressible effects and  mixing layer growth}
As shown in the previous section, effects induced by the adiabatic
gradient start to appear when the mixing length becomes of the order
of the adiabatic length $L_{ml}(t) \sim L_{ad}$. It is nevertheless
possible to study the limit $L(t) \ll L_{ad}$ but still observing
important effects due to compressibility.  Indeed, compressibility due
to stratification is controlled by the Atwood number.  From the
expression of the instantaneous Rayleigh number (\ref{eq:Rat}) one may
compute the typical length scale at which turbulence will be
maximal, i.e. the largest extension of the mixing layer up to which
the Rayleigh number is still growing, before decreasing because of the
adiabatic gradient. This is just given by the maximum of $\tilde
Ra(t)$ as a function of time, which is reached at a characteristic
time, $t^*$ such that: \be L_{ml}(t^*) = \frac{3}{4} L_{ad} =
\frac{3\, c_p\, \Delta T}{4\, g}.  \ee It is also  possible  to estimate the
typical Mach number reached at the maximal turbulent intensity,
considering that hydrodynamical velocities can be estimated as,
$V_{max} \sim d/dt L_{ml}(t^*) = 2 \alpha^{(L)} At\, g\, t^* $ and that the
minimal sound speed is given, in our units, by $v_s = \sqrt{T_d}$, we
get for the Mach number at the maximal turbulent intensity: $Ma \sim
At \sqrt{\alpha^{(L)} c_p}$ where we have used (\ref{eq:Lt5}) to estimate
$t^*$ at a given $L_{ml}(t^*)$. As a result, dynamical compressibility
is only driven by the Atwood number -at fixed $c_p$. Using the typical
values of $\alpha^{(L)} \sim 5\;10^{-2}$, as reported in the
literature \cite{rt4}, and
plugging the correct prefactor, we estimate $Ma \sim 0.4 $, for the
largest Atwood we could achieve, $At \sim 0.4$.\\
It is well known that compressibility effects break the up/down symmetry in the mixing layer propagation \cite{cook,cook2}, downwards spikes (cold fluid blobs) move faster than upwards bubbles (hot fluid blobs). Such effect is completely missing in Boussinesq approximation where there is a perfect up/down symmetry, by definition. 

\begin{figure}
\begin{center}%
\advance\leftskip-0.55cm
\includegraphics[scale=0.7]{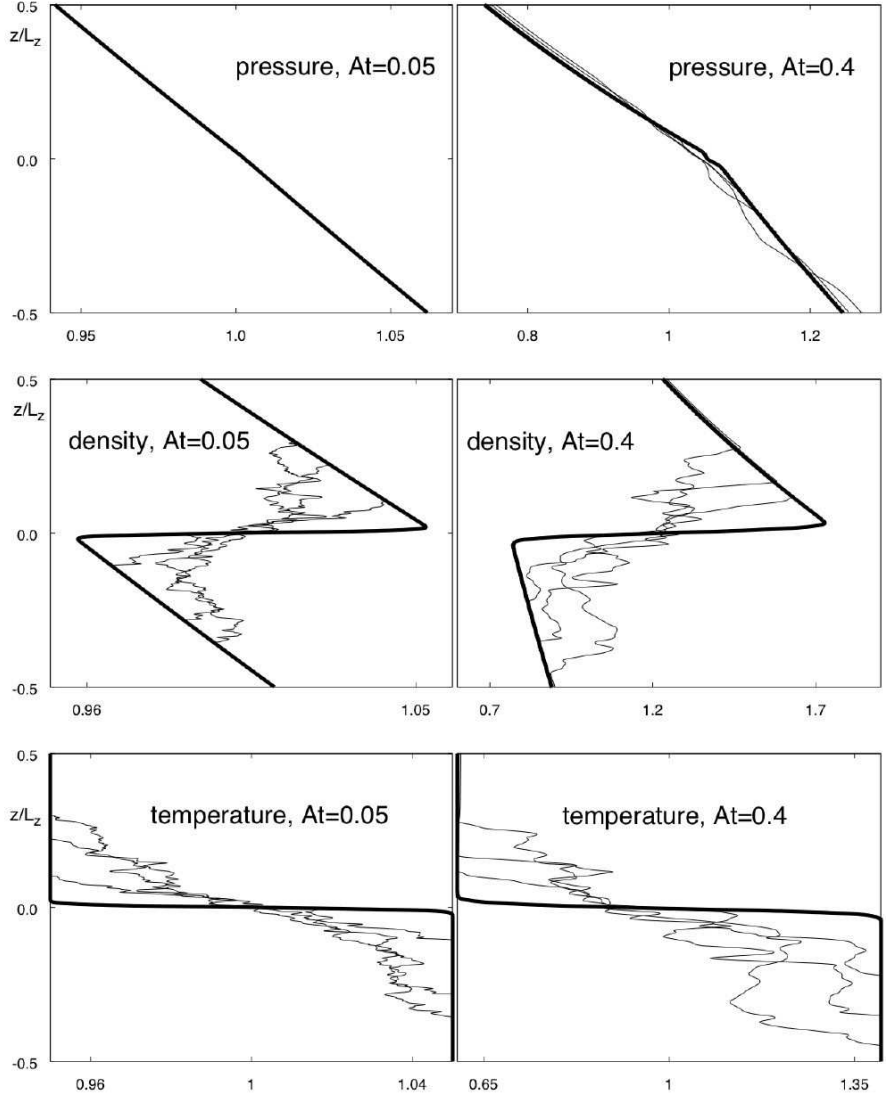}
\caption{Temperature, $\langle T(x,z;t)\rangle_x$,
 density, $\langle \rho(x,z;t)\rangle_x$, and pressure, $\langle p(x,z;t)\rangle_x$, 
instantaneous mean profiles at different time during the RT evolution. 
Left column: $At=0.05$, times $t=3 \tilde{\tau},6 \tilde{\tau},7 \tilde{\tau}$ (run A, table I); Right column: $At=0.4$, times $t=3 \tilde{\tau},4.5 \tilde{\tau},6 \tilde{\tau}$ (run B, table I). Initial hydrostatic profiles are depicted by solid bold lines. Notice the asymmetry for the mixing layer growth in the latter case. Notice also the appearance for high Atwood of hydrodynamical pressure fluctuations superposed to the hydrostatic pressure profiles. Both effects are absent in the small Atwood case.}
\label{fig:profiles}
\end{center}
\end{figure}

Neglecting slowing down effects induced by the adiabatic gradients,
i.e. limiting the study of the mixing layer growth up to $L_{ml}(t)
\ll L_{ad}$, we may investigate the symmetry breaking in our set up at
changing the Atwood number. To give an idea of the effects of
compressibility, we show in figure \ref{fig:profiles} a few
instantaneous mean profile of temperature, density and pressure for 
the two Atwood numbers here investigated. From the density and
temperature profiles it is easy detectable, already by naked eyes, the
asymmetry present for the high Atwood case $At=0.4$ in the growth of
the mixing layer, with the colder and denser front moving
faster. Also, the appearance of non-trivial fluctuations in the
pressure around the hydrostatic profile, for the case at $At=0.4$, are
the clear evidence of compressible effects at play. Both the asymmetry
and the pressure fluctuations are completely absent for the case at
small Atwood (left panels of figure \ref{fig:profiles}, an evidence of
Boussinesq-like thermal fluctuations).  All numerical experiments have
been performed by preparing the initial configuration in its
hydrostatic equilibrium (\ref{eq:hydroRT}) plus a smooth interpolation
between the two half volumes in order to have a finite width of the
initial interface. The initial temperature profile is therefore chosen
to be:
$$ T_0(z) = \frac{T_u+T_d}{2} + \frac{T_u-T_d}{2}  \tanh\left ( \frac{(z-z_c)}{w} \right)$$
where with $w$ we define the initial width of the interface and $z_c$ its unperturbed height ($z_c=0$ in our frame of reference). Initial density $\rho_0(z)$ and pressure $p_0(z)$ are then fixed by solving the hydrostatic equation (\ref{eq:phydroRT}) in order to get the 
 hydrostatic 
 solution corresponding to the smoothed temperature profile. 

To destabilize the initial configuration, we follow \cite{rt.temp} and shift randomly the center of the interface by adding horizontal perturbation at different wavelengths in the range $k \in [k_{min}:k_{max}]$:
\be
z_c \rightarrow z_c(x) = \frac{\epsilon}{N}  \sum_{k=k_{min}}^{k=k_{max}} cos(2\pi\,k\,x/L_x + \phi_k)
\ee
where $\phi_k$ are random phases and $N=\sqrt{k_{max}-k_{min}}$, in order to have a total amplitude for the initial width almost independent on the number of modes. We have tried different ranges of
wavelengths, without observing quantitative differences in the large
time growth of the mixing layer. The ratio $W=\epsilon/w$
gives the ``wiggling'' of the interface, i.e. how much the
perturbation of the interface position  is important with respect to
the interface width.\\ Below, we present results in different
geometry, up to a resolution of $L_x \times L_z= 1664 \times 4400$ with 
 different choices of $W$. For each parameters  set  we made typically ${\cal O}(50)$ separate  RT evolution, starting from different random phases initial configurations. \\

 In the sequel,  we show a summary of the results from two typical numerical series of runs, one with $At =0.05$ (small compressibility) and a second one with $At =0.4$ (large compressibility).
It is useful to adopt a different definition for the  mixing length in terms of a 
 bulk mixing percentage, introducing the characteristic function  (tent-map):
\be
\label{eq:mL}
\begin{cases}
\chi[\xi] = 2 \xi;\qquad 0 \le \xi \le 1/2 \\
\chi[\xi] = 2\,(1-\xi);\qquad 1/2 \le \xi \le 1 \\
\end{cases}
\ee
and defining the mixing length as \cite{cook2}: 
\be
H(t) = \frac{1}{L_x} \int dx dz\, \chi \left[ \frac{T(x,z)-T_u}{T_d-T_u} \right]. 
\ee
It is easy to realize that if the temperature is fully homogenized in the fluid, $T(x,z) = (T_u+T_d)/2$, 
then the mixing length coincides with the full vertical extension of the box: $H = L_z$; if we have two perfectly separated hot and cold regions we have $H=0$. In the intermediate situation when we have a mean linear temperature profile for $z \in [z_d,z_u]$, between two unmixed regions ($T=T_u$ if $z>z_u$  and 
$T=T_d$ if $z<z_d$)  the mixing length estimated by (\ref{eq:mL}) is exactly given by half of the linear region, $H= (z_u-z_d)/2.$. The definition of the mixing length (\ref{eq:mL})  must be preferred with respect to 
more common definition of $L_{ml}$  based on thresholds on the linear profile, as adopted in the previous 
section.  The former, being based on  a bulk measure is not affected
too much on the highly fluctuating properties of the interface between
mixed and unmixed fluids. This is particularly important in 2d, where
the averaged profile, being a  one-dimensional cut, may 
fluctuate a lot (see also figure \ref{fig:profiles}). Anyhow,
in the case of a perfectly linear temperature profile the two lengths are obviously related by the relation
 $H = 1/2 \delta L_{ml}$, where $\delta$ is the percentage threshold 
used to identify the mixing front (in the previous section $\delta =0.99$). 

Moreover, because here we want to distinguish the downward growth of the front due to  
cold spikes from the upward growth of bubbles, we introduce two different integral mixing lengths:
\begin{eqnarray}
  H_s(t) = \frac{1}{L_x} \int dx dz\, \left(\Theta\left(\frac{L}{2}-z\right) \chi \left[ \frac{T(x,z)-T_u}{T_d-T_u} \right]\right); \nonumber \\
  H_b(t) = \frac{1}{L_x} \int dx dz\, \left(\Theta\left(z-\frac{L}{2}\right) \chi \left[\frac{T(x,z)-T_u}{T_d-T_u} \right]\right); \nonumber
\end{eqnarray}
where 
of course, $H(t) = H_s(t)+H_b(t)$.  Clearly, the $\alpha^{(H)}$ value ruling
the long term quadratic growth of the integral mixing $H$ is not
necessarily the same of $L_{ml}$.  Typically one expects the same
relation $\alpha^{(H)} = 0.5\, \delta\, \alpha^{(L)}$ valid for the definition
of the two mixing length, at least for times long enough.

\begin{figure}
\advance\leftskip-0.65cm
\includegraphics[width=0.5\textwidth]{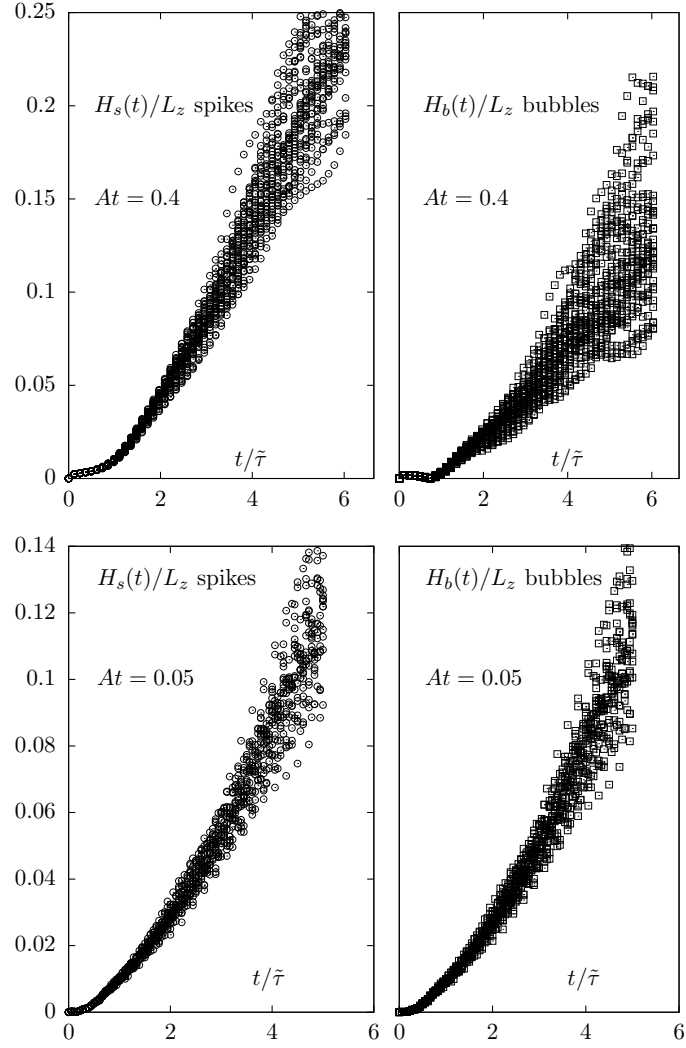}
\caption{Growth of $H_{s,b}(t)$, run-by-run, for the two Atwood numbers of run A and run B.  The mixing length width is normalized by the total box width, $L_z$. Notice the evident asymmetry between spikes and bubbles for the high Atwood case (run B).}
\label{fig:alpha1}
\end{figure}

\begin{figure}
\begin{center}
\advance\leftskip-0.55cm
\includegraphics[width=0.5\textwidth]{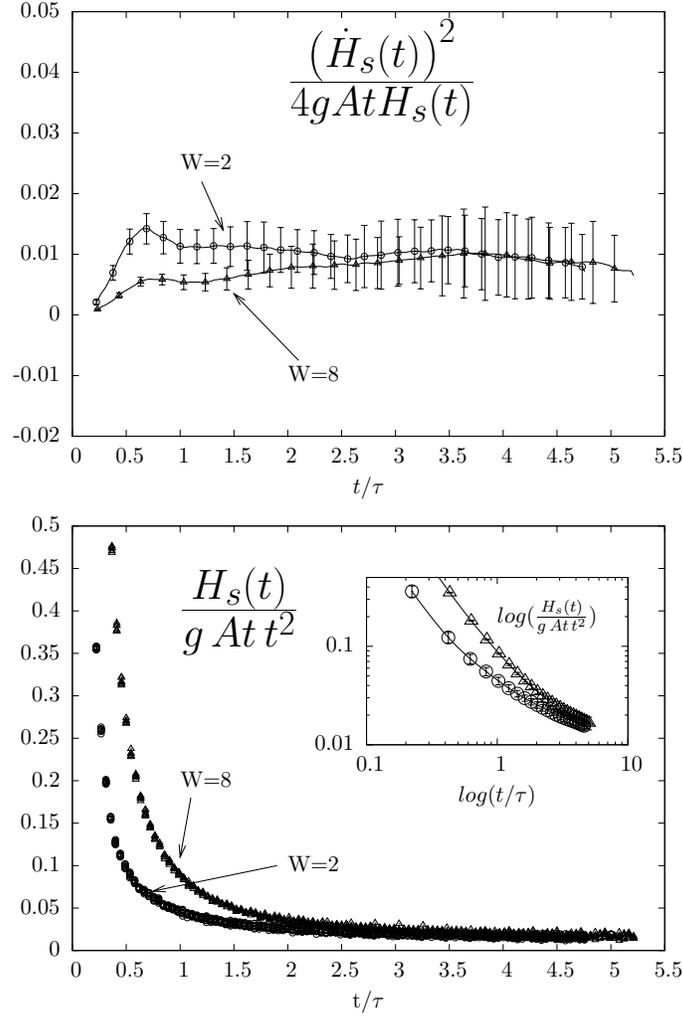}
\caption{Run $A$. $At=0.05$.  Analysis of the asymptotic growth rate for
  spikes, $\alpha^{(H)}_s$.  Bottom panel: mixing length evolution
  normalized by $t^2$ for two different set of initial width, $W =
  \epsilon/\omega = 2$ (circles) and $W=8$ (triangles), where
  $\epsilon$ is the intensity of the initial perturbation and $\omega$
  is the width of the regularizing $tanh$ initial profile. Data refers
  to $N_{conf}=50$ for both cases. Notice the long relaxation time
  before the two evolution forgets the initial conditions. This is due
  to the presence of the prefactor proportional to $L_{ml}(t_0)$ in 
the sub-leading linear term of eq. (\ref{eq:Lt5}). 
Inset: the
  mean value of the data shown in
  the body but in log coordinates -same symbols.\\
  Upper panel: mean value of the instantaneous growth rate of spikes
  extracted from (\ref{eq:histo}) for the two initial set up with
  $W=2,8$.  Average is performed over $N_{conf}=50$ separate Rayleigh
  Taylor evolution for the two cases.  Error bars are estimated out of
  root mean square fluctuations. Notice the more extended range where
  the two set-up superpose and the extended time interval where
  $\alpha^{(H)}_s$ stays constant (notice the different y-scale between
  lower and upper panels). Results for bubbles evolution are similar
  and not shown. Both cases are summarized in figure \ref{fig:histo}}
\label{fig:alpha2}
\end{center}
\end{figure}

As one can see in figure \ref{fig:alpha1} there is a wide scattering of
the mixing length evolution from run to run, where the only
differences between them is the realization of the initial random
phases. 
Due to the intense local temperature and density fluctuations, averaging over horizontal direction is not very efficient to smooth down statistical fluctuations, and one observes high variations from sample to sample:
 many realizations are needed to
extract stable quantitative results on the long time evolution. 
In order to have an insight on the typical fluctuations we decided to
analyze run by run and following two fitting procedures.  First, we  start from the equivalent of (\ref{eq:Lt5}), written for bubbles
and spikes separately: 
\be
\begin{cases}
 H_{b}(t-t_0) = H_{b}(0) + \dot H_{b}(0)\, t  + \alpha^{(H)}_{b}\, At\,g\, t^2  \\
H_{s}(t-t_0) = H_{s}(0) + \dot H_{s}(0)\, t +
\alpha^{(H)}_{s}\, At\,g\, t^2 \\
\end{cases}
\label{eq:H}
\ee
with $\dot H_{b,s}(0) =  2\,\sqrt{H_{b,s}(0) \, \alpha^{(H)}_{b,s}\, At\, g}$,
where $t_0$ must be understood as the time when the initial
perturbation is fully entered in its non-linear regime. In other
words, $t_0$ must be larger than the typical characteristic time of
the slowest unstable mode. It can be estimated from linear stability
analysis as $t_0 \sim \sqrt{L_x/(2 \pi \,g \, At)} $. A brute force way
to extract the growth rate is to evaluate the ratio $\alpha^{(H)}_{s,b} =
\lim_{t\rightarrow \infty} H_{s,b}(t)/t^2$. Even, neglecting possible
contamination due to stratification, this is of course valid, only
asymptotically, when both dependencies on the initial time $t_0$ and on
the initial mixing length $H_{s,b}(t_0)$ become negligible. As a
matter of fact, taking into account also the maximum time
achievable due to numerical limitations, it is very difficult to
extract stable statistical results on the $\alpha^{(H)}$ fluctuations
starting from the brute force analysis of (\ref{eq:H}). For instance, we
found that a parabolic fit to our data, taking $\alpha^{(H)}_{s,b}$ free is
very sensitive to the initial time $t_0$ and/or the initial distance
$H_{b,s}(t_0)$, without allowing for a systematic assessment of the
asymptotic behaviour.  To give an idea of the importance of the
initial condition versus statistical fluctuations, we show in the
bottom panel of figure \ref{fig:alpha2} the results of the asymptotic
ratio $H_{s,b}(t)/t^2$ for two different series of runs with different
initial conditions. As one can see, even if asymptotically there is a
clear tendency to forget the initial separation, in agreement with
(\ref{eq:H}) there is not
a well developed plateau, up to the time achievable in out numerics, indicating the existence of important
sub-leading effects. The existence of such terms is highlighted in the
inset of the same panel, where a log-log plot clearly shows the lack
of a plateau even for large times.\\
Another alternative, and more robust way, to extract $\alpha^{(H)}$ relies
on the differential equivalent of (\ref{eq:H}) given by
(\ref{eq:cook}) or (\ref{eq:luca-andrea}) when stratification becomes
important. Using (\ref{eq:cook}), one may directly assess the
non-linear growth rate, without spurious contamination from initial
conditions.

\begin{figure}
\begin{center}
\advance\leftskip-0.55cm
\includegraphics[width=0.6\textwidth]{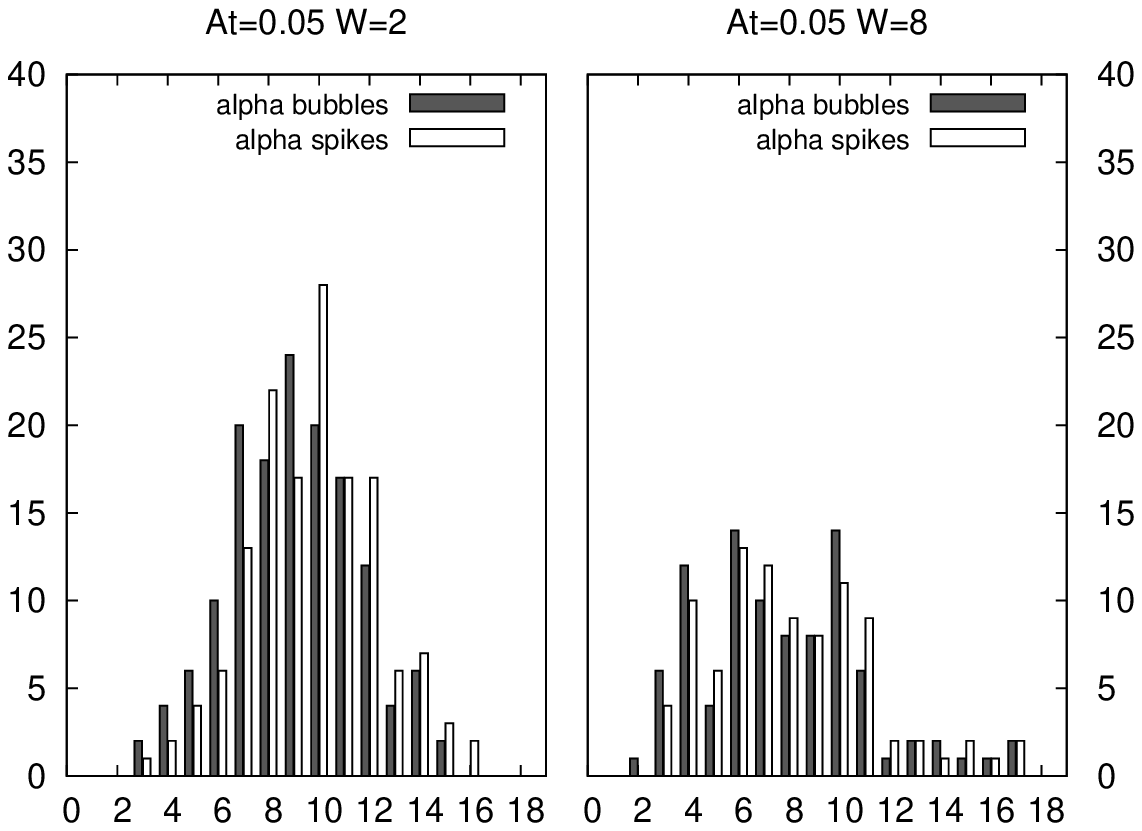}
\includegraphics[width=0.6\textwidth]{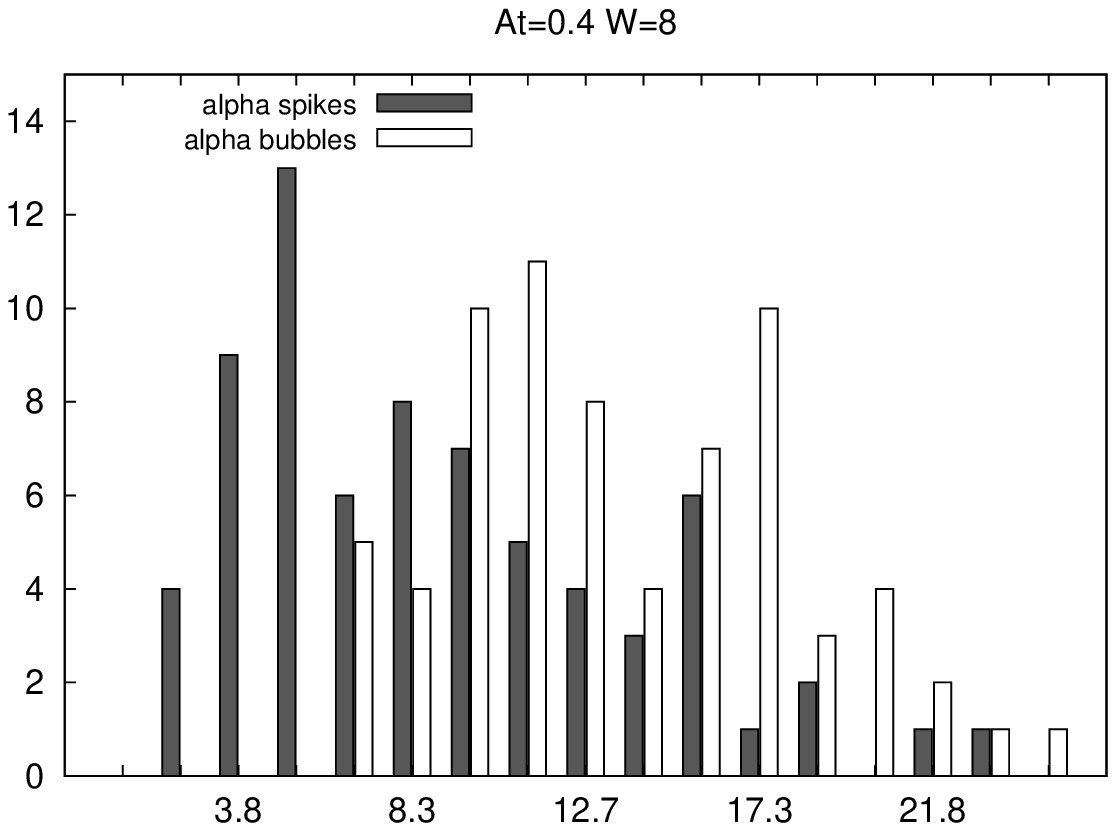}
\caption{Top: Run $A$, $At=0.05$;  histograms of $\alpha^{(H)}_{s,b}$ (multiplied by $10^3$ for the sake of clarity) as extracted
from (\ref{eq:histo}), 
at fixed initial width $W=2$ (left) and $W=8$ (right).
 The fit is done over $50$ and $35$
  different configurations respectively. In order to test dependency on 
the fitting window we have summed results from
 two different ranges, $ t \in [1.5 \tilde{\tau}:4.5 \tilde{\tau}]$ and $ t \in [2.2 \tilde{\tau}:4 \tilde{\tau}]$  
in both cases the maximum time is such that the front didn't reach more than $80\%$ of the total vertical 
extension of the physical domain. Bottom: Run $B$, $At=0.4$. Results
from two fitting ranges $ t \in [2.3 \tilde{\tau}:5.4 \tilde{\tau}]$ and 
$ t \in [3 \tilde{\tau}:4.5 \tilde{\tau}]$.  Notice the asymmetry developing for $At=0.4$, with spikes traveling faster. 
An estimate of the mean value for the growth rate 
in the two cases gives: $\alpha^{(H)}_s = (10\pm2)\;10^{-3} $ and 
$\alpha^{(H)}_b = (9.5\pm2)\;10^{-3} $ at $At=0.05$,
while $\alpha^{(H)}_s = (14\pm4)\; 10^{-3} $ and  $\alpha^{(H)}_b = (9\pm5)\;10^{-3} $ for $At=0.4$.} 
\label{fig:histo}
\end{center}
\end{figure}

In the upper panel of figure \ref{fig:alpha2} we show the same data
plotted in the lower panel but for the ratio \be \alpha^{(H)}_{s,b} = (\dot
H_{s,b}(t))^2/ (4 g\, At\, H_{s,b}(t)),
\label{eq:histo}
\ee i.e. we address time-by-time the part depending on asymptotic
growth rate only. It is evident the net improvement in both the
extension of the range where $\alpha^{(H)}$ coefficients are constant and
the clear disentanglement of effects coming from the initial
conditions.  Out of the data for $\left(\dot H_{s,b}(t)\right)^2/4
\left(g\, At\, H_{s,b}(t)\right) $ we may estimate the statistical fluctuations of
$\alpha^{(H)}_{s,b}$, by making a fit to a constant in a given time windows.
In figure \ref{fig:histo} we plot the results of fitting the evolution
(\ref{eq:histo}) independently for bubbles or spikes (upward or
downward fronts). From this we learn a few interesting facts: (i) at
small Atwood (upper panel) bubbles and spikes travels almost with the
same statistics, even though a small asymmetry can be observed in the
shape of the whole histogram. The asymmetry is so small, that if
averaged quantities are measured, the differences between them falls
within error bars; (ii) there are not important effects form initial
conditions -compare the two upper  panels  obtained with two
different classes of initial conditions-; at least when data are
fitted using (\ref{eq:histo}), confirming that the observed
spatio-temporal evolutions is dominated by strongly non-linear fully
developed dynamic; (iii) at large Atwood (lower panel) the asymmetry becomes
evident, spikes are systematically faster then bubbles, the two
evolutions gives different mean vales for $\alpha^{(H)}_s$ and $\alpha^{(H)}_b$ parameter. Our measure of the average
global growth rate $\alpha^{(H)}$, can  be estimated by summing up
the growth rate in the two half cells: 
$\alpha^{(H)}=\alpha^{(H)}_s+\alpha^{(H)}_b \sim 0.02$
is agreement with values typically found in literature 
\cite{rt4,cook,cook2}. For instance, in \cite{rt4} 
a detailed overview of numerical results gives for the growth rate of
bubbles,   measured on the $99\%$
width, $\alpha^{(L)}_b \sim 0.025 \pm 0.003$, in agreement with
$\alpha_b^{(H)} = 0.0095 \pm 0.002$ we found for our integral growth
rate (see caption of figure \ref{fig:histo})  
taking into account that by definition one expects a factor two
between the measurement made on the integral quantity, $\alpha^({H})$,
 and the
measurement made on the $99 \%$ level set, $\alpha^{(L)}$.

\begin{figure}
\begin{center}
  \advance\leftskip-0.55cm
  \includegraphics[width=0.7\textwidth]{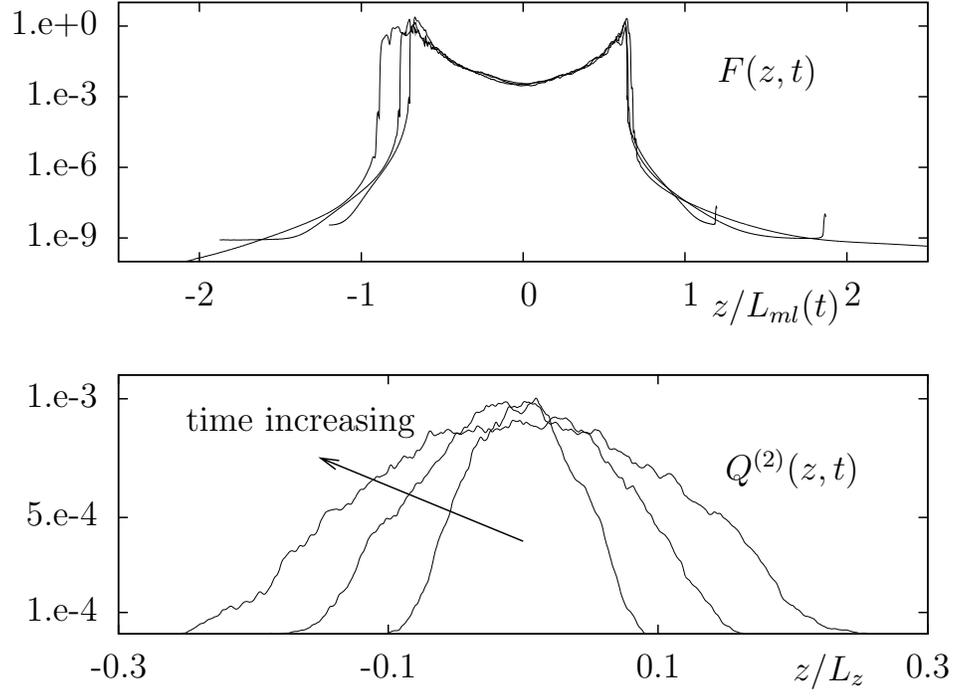}
  \caption{Bottom: 2nd order moment of temperature fluctuations (see
    Eqn. (\ref{eq:Tmom})) as a function of the height at times
    $t=(2.2 ,3.3 ,4.4) \tilde \tau$. The
     $z$-height has been normalized with the total cell extension $L_z$. Top: Flatness 
$Q^{(4)}(z,t)/(Q^{(2)}(z,t))^2$
    at the same three instants of time as in the bottom panel. The
     $z$-height has been normalized with $L_{ml}(t)$ in
    order to show the self-similarity of the mixing process (the three
    curves collapse onto each other by rescaling). Parameters refer to
    run A in table 1.}
  \label{fig:mixing}
\end{center}
\end{figure}

The last issue we want to discuss concerns with homogenization inside
the mixing layer. It is easy to show that in the Boussinesq
approximation for a convective stationary cell with a mean linear
temperature profile, all deviations from the mean profiles are
homogeneous. The case of RT evolutions investigated here is slightly
different. First, whenever stratification is important, there is no
reason to expect exactly homogenization inside the mixing
length. Second, and more importantly, homogeneity must be expected
only well inside the mixing layer, far from the up and downside
fronts, where clearly strong non-homogeneous effects for both mean and
fluctuating quantities must appear. It is interesting therefore to
test, how homogeneous the statistics is, also to quantify the degree
of mixing.  In order to do that, we introduce the $p$-th order moments
of temperature fluctuations: 
\be\label{eq:Tmom} Q^{(p)}(z,t) = \langle
(T(x,z) - \langle T(x,z)\rangle_x)^p \rangle_x.  \ee In
 figure \ref{fig:mixing} we show the root mean square fluctuations around
the vertical mean temperature profile, $Q^{(2)}(z,t)$, (bottom panel)
and the flatness, $F(z,t) = Q^{(4)}(z,t)/(Q^{(2)}(z,t))^2$,
i.e. the ratio between fourth and squared second order moments of
fluctuating quantities (top panel). As one can see, the root mean square
fluctuations tend --very slowly-- to develop a flatter and flatter
plateau inside the mixing region, demonstrating that if the mixing
layer is wide enough, there will be a larger and larger region where
statistics is pretty homogeneous.  On the other hand, if we plot the
Flatness as a function of a normalized mixing length width, it
converges towards a self-similar profile, for any time, where the
effects coming from the two boundaries of the mixing regions are felt
inside the whole layer, without showing any trend towards
homogenization.  This second finding is a clear indication that
if normalized with the total mixing length extension, the region where
the statistics may be considered homogeneous does not increase with
time.

\section{Conclusions and perspectives} 
\label{sec:conclusions}
We have explicitly computed the continuum thermo-hydrodynamical limit
of a new formulation of Lattice Kinetic equations for thermal
compressible flows, recently proposed in \cite{JFM.nostro} We have
shown that the hydrodynamical manifold is given by the correct
compressible Fourier-Navier-Stokes equations for a perfect fluid. We
have validated the calculations against exact results for transition
to convection in \RB\ compressible systems and against direct
comparison with finite-difference methods. The method is stable and
quantitatively reliable up to temperature jumps between top and bottom
walls (stratification) of the order of $\Delta T/ T_u \sim 2$.
We have also applied the method to study \RT\ instability for
compressible stratified flows and we determined the growth of the
{\it asymmetric}
 mixing layer at changing Atwood numbers up to $At \sim 0.4$ and to
Rayleigh $Ra \sim 2 \times 10^{10}$. We determined the distribution of
the growth rate for bubbles and spikes, at changing $At$ and we
discuss its dependence on the initial perturbation. \\ We also
discussed the importance of the adiabatic gradient for the growth of
the RT mixing layer in strongly stratified systems. In the latter
case, we showed the existence of a maximal width, the adiabatic
length, $L_{ad}$, for the mixing region.  The high flexibility --and
locality-- of LB algorithm makes them the ideal playground where to
push the resolution, having perfectly scalable performances as a
function of the number of processors in the parallel architecture. In
particular, it is simple to extend such algorithm to deal with fully
3d systems for ideal, non-ideal and/or even immiscible
two fluids systems. High resolution studies of  \RT\ systems meant to
investigate short wavelengths scaling properties of 
velocity, density and temperature
fields for high Rayleigh, with and without surface tension
\cite{mazzino}, and  using a highly optimized LBM  algorithm for
the Cell Broadband Engine \cite{lele1}
 are under current investigation and 
 will be reported elsewhere \cite{lele}.  
 The thermal LBM
here proposed still suffers of small spurious oscillations of
temperature and perpendicular velocity close to the solid boundaries,
making it still not appropriate to study high Rayleigh numbers
stationary convection.  A possible way to overcome this difficulty
consists in abandoning numerical schemes based on exact streaming and
to develop the proposed thermal LBM on a finite volume scheme. Results
in this direction are out of the scope of this paper and will be the
subject of a forthcoming publications.

{\sf Acknowledgments}: We thank R. Benzi, G. Boffetta, A. Celani,
H. Chen, S. Gauthier, A. Mazzino  and
X. Shan  for useful discussions.
We acknowledge partial computational support from ``Centro Enrico
Fermi'', from CASPUR (Roma, Italy, HPC grant n. std09-327), from
CINECA (Bologna, Italy) and SARA (Amsterdam, The Netherlands).  This
work was carried out under the HPC-EUROPA2 project (project number:
228398) with the support of the European Commission Capacities Area -
Research Infrastructures Initiative. AS thanks FT and the Technical
University of Eindhoven for hospitality during his HPC-EUROPA visit.

\section{APPENDIX A}

\begin{figure}
\begin{center}
\includegraphics[scale=1.]{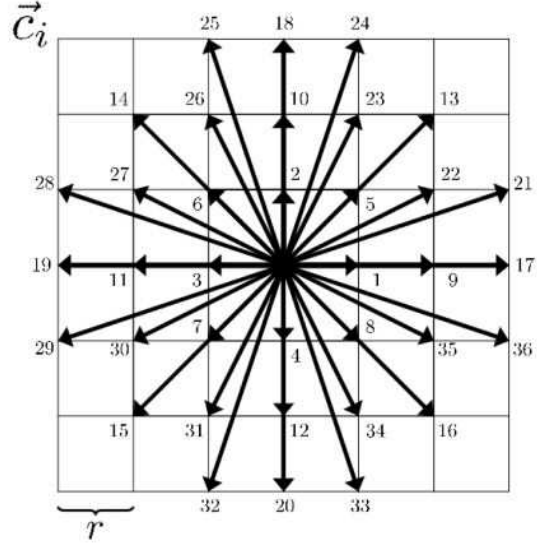}
\caption{Scheme for the $D2Q37$ model used for the simulation of thermohydrdoynamics. The 'lattice constant' is $r \sim 1.1969$ as reported in \cite{Shan07}. The velocity set is such that every projection of the velocity is an integer multiple of $r$ which is chosen to enforce the unitarity of Hermite polynomials (\ref{H1},\ref{H2}) up to the fourth order. The relationship between real and velocity lattice is set by $\Delta x= r \Delta t$ with $\Delta x$ and $\Delta t$ space and time discretizations. Based on the Hermite-Gauss quadrature procedure  \cite{Shan06,Shan07,Philippi06}, the $D2Q37$ can be regarded as the minimal  on grid square lattice giving with accurate Hermite polynomials up to the fourth order.   This quadrature  ensures that the Navier-Stokes thermodynamics is recovered with full Galilean invariance. Lattice $D2V37$ firstly appeared and was shown to be minimal for $2d$ fourth order models  in Reference \cite{Philippi06}, where the authors formally showed the equivalence between the condition of norm preservation and the preservation of the orthogonality property, in constructing these sets of lattice vectors.}
\label{fig:37}
\end{center}
\end{figure}

In this appendix we detail the steps of the Chapman Enskog expansion
leading to the thermohydrodynamical equations under the effect of
general forcing term $\rho {\bm g}$. Similar analysis (without the
effect of the forcing) can be found in  \cite{Siebert07}. We start from the shifted equilibrium
formulation \be
\label{MAIN}
f_{l}({{\bm x}+{\bm c}_l}\Delta t,t+\Delta t)-f_{l}({\bm
  x},t)=-\frac{\Delta t}{\tau} \left(f_{l}({\bm x},t)-\bar f_{l}\
\right) \ee where, for the sake of simplicity, in the notation of this
appendix we have renamed the equilibrium distribution function with
shifted fields, $f^{(eq)}_l = \bar f_l$:
$$\bar f_{l}=\bar f_{l}(\rho,{\bm u}^{(L)}+{\bm \chi},T^{(L)}+ \lambda)$$
and where ${\bm \chi}$ and $\lambda$ are general momentum and
temperature shifts for the equilibrium distribution with ${\bm
  u}^{(L)}$,$T^{(L)}$ the lattice velocity and temperature hereafter
denoted simply with ${\bm u}$ and $T$. Central to our analysis is the
expansion of the equilibrium distribution in Hermite polynomials
\cite{Philippi06,Shan06,Siebert07}
$$\bar f_{l}=w_l \sum_{n} \frac{1}{n!} {\bm a}^{(n)}_0 (\rho,{\bm u}+\epsilon {\bm \chi},T+\epsilon^2 \lambda) {\cal H}^{\
  (n)}_{l}$$ with $w_l$ suitable weights whose values are reported in
\cite{Shan07,Philippi06} for the $D2Q37$ model here used (see also figure
\ref{fig:37}).  For the purposes of our investigation a fourth order
approximation proves to be enough to recover the correct equations
with the right isotropic properties for all hydrodynamical fields and
tensors up to the eighth order \cite{Shan07}. The Hermite polynomials
are given by the following relations: \be\label{H1} {\cal
  H}^{(0)}_l=1; \hspace{.2in} {\cal H}^{(1)}_{l}={\bm c}_l;
\hspace{.2in} {\cal H}^{(2)}_{l}={\bm c}_l^2-{\bm \delta} \ee
\be\label{H2} {\cal H}^{(3)}_{l}={\bm c}_l^3-{\bm \delta} {\bm
  c}_{l};\hspace{.2in} {\cal H}^{(4)}_{l}={\bm c}_l^4-{\bm \delta}
{\bm c}_{l}^2+{\bm \delta} {\bm \delta} \ee and the projection
coefficients ${\bm a}_{0}^{(n)}$ by
$$
\begin{cases}
{\bm a}^{(0)}_0=\rho \vspace{0.1in} \\
{\bm a}^{(1)}_0=\rho {\bm u}+ \epsilon \rho {\bm \chi}  \vspace{0.1in}\\
{\bm a}^{(2)}_0=\rho [{\bm u}^2+(T-1){\bm \delta} ]+ \epsilon \rho {\bm \chi} {\bm u}+\epsilon^2 ( \rho {\bm \chi}^2 \\
\hspace{.3in} +\rho {\bm \delta} \lambda ) \vspace{0.1in}\\
{\bm a}^{(3)}_0 =\rho [{\bm u}^3+(T-1){\bm \delta} {\bm u} ]+ \epsilon (\rho {\bm \chi} {\bm u}^2 +\rho (T-1){\bm \delta} 
{\bm \chi})\\
\hspace{.3in} +\epsilon^2 ( \rho {\bm \chi}^2{\bm u}  +\rho \lambda {\bm \delta} {\bm u} ) +\epsilon^3 (\rho  {\bm \chi\
}^3+\rho \lambda {\bm \delta} {\bm \chi}  )  \vspace{0.1in}\\
{\bm a}^{(4)}_0=\rho [{\bm u}^4+(T-1){\bm \delta} {\bm u}^2+(T-1)^2{\bm \delta}^2]\\
\hspace{.3in} +\epsilon (\rho {\bm \chi} {\bm u}^3 +\rho (T-1){\bm \delta} {\bm \chi} {\bm u})\\
\hspace{.3in} +\epsilon^2 ( \rho {\bm \chi}^2{\bm u}^2  +\rho \lambda {\bm \delta} {\bm u}^2 +\rho (T -1) {\bm \delta} \
{\bm \chi}^2) \\
\hspace{.3in}  + \epsilon^3 (\rho  {\bm \chi}^3{\bm u}+\rho \lambda {\bm \delta} {\bm \chi}{\bm u})+ \epsilon^4  (\rho \
{\bm \chi}^4 +\rho \lambda {\bm \delta} {\bm \chi}^2)
\end{cases}
$$
where the shorthand notations of Grad \cite{Grad49,Shan06} for fully
symmetric tensors are adopted.  A possible set of {\it on-site}
space-filling lattice velocities can be found in figure \ref{fig:37}
and fully detailed in \cite{Shan06,Philippi06,Surmas09}. If one gives
up the requests to have lattice velocities only on grid points and
allows also for out of lattice discretized velocity sets, the number
of vectors needed to recover isotropy for moments up to order eight
can be reduced \cite{Surmas09}.  We next introduce \cite{Buick00} a
small separation of scale parameter $\epsilon$ and consider the
expansion in $\epsilon$ for the distribution function 
\be\label{EX1}
f_l=f_l^{(0)}+\epsilon f_l^{(1)}+\epsilon^2 f_l^{(2)}+\epsilon^3
f_l^{(3)}+\epsilon^4 f_l^{(4)}+.....  \ee and the rescaling of the
time-space derivatives \be\label{EX2}
\partial_{t} \rightarrow
\epsilon \partial_{1t}+\epsilon^2 \partial_{2t}+{\cal O}(\epsilon^3);
\hspace{.2in} \partial_{i} \rightarrow \epsilon \partial_{i}.  \ee
This allows to rewrite the streaming term in the lattice Boltzmann
equation as
$$
f_{l}({{\bm x}+{\bm c}_l \Delta t},t+\Delta t)-f_{l}({\bm x},t)= \epsilon A_1+\epsilon^2 A_2+\epsilon^3 A_3+...
$$
where for our purposes it is enough to consider terms up to $A_2$
$$
\begin{cases}
A_1= (\partial_{1t} f^{(0)}_l+c^i_{l} \partial_{1t} \partial_{i} f^{(0)}_l)\Delta t \vspace{0.1in}\\
A_2= (\partial_{2t} f^{(0)}_l+\partial_{1t} f^{(1)}_l+c_l^i \partial_{i} f^{(1)}_l)\Delta t  +\frac{1}{2}(c_l^i c_l^j \partial_{i} \partial_{j} f^{(0)}_l  \\
\hspace{.3in} +c_l^i \partial_i \partial_{1t} f^{(0)}_l+ c_l^i \partial_i \partial_{1t} f^{(0)}_l+\partial_{1t} 
\partial_{1t} f^{(0)}_l)\Delta t^2.
\end{cases}
$$
If we further rescale the shifting \cite{Buick00} fields as
\be\label{SHIFTresc} {\bm u} \rightarrow {\bm u} + \epsilon {\bm
  \chi}; \hspace{.2in} T \rightarrow T + \epsilon^2 \lambda \ee the
shifted equilibrium can be further seen as a power series in
$\epsilon$
$$                                                                                                                      
\bar f_{l}(\rho,{\bm u}+\epsilon {\bm \chi},T+\epsilon^2 \lambda)=\bar
f_l^{(0)}+\epsilon \bar f_l^{(1)}+\epsilon^2 \bar f_l^{(2)}+\epsilon^3
\bar f_l^{(3)}+\epsilon^4 \bar f_l^{(4)}+....
$$
with
$$
\begin{cases}
  \frac{\bar f^{(0)}_l}{w_l}=\rho {\cal H}_l^{(0)}+ \rho {\bm u} {\cal
    H}_l^{(1)}+\frac{1}{2}\rho [{\bm u}^2+(T-1)
  {\bm \delta} ]{\cal H}_l^{(2)} \\
  \hspace{.3in} +\frac{1}{6}\rho [{\bm u}^3+(T-1){\bm \delta} {\bm u} ]{\cal H}_l^{(3)}  \\
  \hspace{.3in} +\frac{1}{24} \rho [{\bm u}^4+(T-1){\bm \delta} {\bm
    u}^2+(T-1)^2{\bm \delta}^2] {\cal H}_l^{(4)}
  \vspace{0.1in} \\
  \frac{\bar f^{(1)}_l}{w_l}=\rho {\bm \chi} {\cal H}_l^{(1)}+ \frac{1}{2} \rho {\bm \chi} {\bm u} {\cal H}_l^{(2)} \\
  \hspace{.3in} +\frac{1}{6} (\rho {\bm \chi} {\bm u}^2 + \rho (T-1){\bm \delta} {\bm \chi}) {\cal H}_l^{(3)} \\
  \hspace{.3in} +\frac{1}{24}(\rho {\bm \chi} {\bm u}^3 +\rho
  (T-1){\bm \delta} {\bm \chi} {\bm u}){\cal H}_l^{(4)}
  \vspace{0.1in} \\
  \frac{\bar f^{(2)}_l}{w_l}=\frac{1}{2}( \rho {\bm \chi}^2 +\rho {\bm
    \delta} \lambda ) {\cal H}_l^{(2)}+\frac{1}{6}
  (\rho {\bm \chi}^2{\bm u}  +\rho \lambda {\bm \delta} {\bm u} ) {\cal H}_l^{(3)}  \\
  \hspace{.3in} +\frac{1}{24}( \rho {\bm \chi}^2{\bm u}^2 +\rho
  \lambda {\bm \delta} {\bm u}^2 +\rho (T -1) {\bm \delta}\
  {\bm \chi}^2) {\cal H}_l^{(4)} \vspace{0.1in}\\
  \frac{\bar f^{(3)}_l}{w_l}=\frac{1}{6}(\rho {\bm \chi}^3+\rho
  \lambda {\bm \delta} {\bm \chi} ) {\cal H}_l^{(3)}+
  \frac{1}{24}(\rho  {\bm \chi}^3{\bm u}+\rho \lambda {\bm \delta} {\bm \chi}{\bm u}) {\cal H}_l^{(4)} \vspace{0.1in}\\
  \frac{\bar f^{(4)}_l}{w_l}=\frac{1}{24}(\rho {\bm \chi}^4 +\rho
  \lambda {\bm \delta} {\bm \chi}^2) {\cal H}_l^{(4)}
\end{cases}
$$
where, upon dimensional considerations, we have requested that when
the forcing rescales as $\epsilon$, the temperature shifting term is
rescaling like $\epsilon^2$ (see also \cite{Buick00} for a more
detailed discussion). Using the Taylor expansion of $f_{l}({{\bm
    x}+{\bm c}_l}\Delta t,t+\Delta t)$, we can impose the consistency
in (\ref{MAIN}) order \ by order in $\epsilon$: \be \label{eq:ce}
\begin{cases}
  {\cal O}(\epsilon^0):  f^{(0)}_l=\bar{f}^{(0)}_l   \vspace{0.1in} \\
  {\cal O}(\epsilon^1): \partial_{1t} f^{(0)}_l+c^i_{l} \partial_{i}
  f^{(0)}_l=-\frac{1}{\tau}(f^{(1)}_l-\bar{f}^{(1)}_l)
  \vspace{0.1in}\\
  {\cal O}(\epsilon^2): \partial_{2t} f^{(0)}_l+\partial_{1t}
  f^{(1)}_l+c_l^i \partial_{i} f^{(1)}_l+(\frac{1}{2} c_l^i
  c_l^j \partial_{i} \partial_{j} f^{(0)}_l \\
  \hspace{.4in} +\frac{1}{2} c_l^i \partial_i \partial_{1t} f^{(0)}_l
  +\frac{1}{2} c_l^i \partial_i \partial_{1t}
  f^{(0)}_l+\frac{1}{2} \partial_{1t} \partial_{1t} f^{(0)}_l)\Delta t  \\
  \hspace{.4in} =-\frac{1}{\tau}(f^{(2)}_l-\bar{f}^{(2)}_l ).
\end{cases}
\ee
Taking the momenta at the zeroth order in $\epsilon$ we can find
some constraints for the higher terms in the expansion\
in of the distribution function. Since we know that
$f^{(0)}_l=\bar{f}^{(0)}_l$, it follows from the definition of
macroscopic fields that
$$                                                                                                                     
\sum_l f_l^{(n)}=0 \hspace{.2in} \sum_l c_l^i f_l^{(n)}=0
 \hspace{.2in} \sum_l c_l^2 f_{l}^{(n)}=0 \hspace{.2in} n \ge 1.
$$                                                                                                                     
\subsubsection{Zeroth order}\label{sub:A}
At the zeroth order in $\epsilon$ we can find some constraints for the
higher terms in the expansion of the distribution function. We know
that
$$                                                                                                                      
f^{(0)}_l=\bar{f}^{(0)}_l .
$$
It follows that, since we define our macroscopic variables as
$$                                                                                                                      
\rho=\sum_l f_l; \hspace{.1in} \rho u_i =\sum_l f_l c_l^i;
\hspace{.1in} T=\frac{1}{D}\sum_l f_l |{\bm c}_l-{\bm u}|^2,
$$
we immediately recover that \be\label{CONSTRAINT1} \sum_l f_l^{(n)}=
\sum_l c_l^i f_l^{(n)}= \sum_l |{\bm c}_l-{\bm u}|^2 f_l^{(n)}=0
\hspace{.2in} n\ge 1.  \ee The last equation leads to (we take the
convention that double indexes are summed upon)
$$\delta_{ij} \sum_{l} (c^i_l c^j_l+u_i u_j-u_i c_l^j-u_j c_l^i   ) f^{(n)}_l=0  \hspace{.2in} n \ge 1                    
$$
that, combined with the constraints for the momentum ($\sum_l c_l^i
f_l^{(n)}=0$), is equivalent to \be\label{CONSTRAINT2} \sum_l c_l^2
f_{l}^{(n)}=0 \hspace{.2in} n \ge 1.  \ee
\subsubsection{First order}
We first evaluate and also remind the values of some useful quantities
that can be easily obtained knowing the relation between Hermite
polynomials and the velocity set (\ref{H1},\ref{H2}) and also the
constraints coming from (\ref{CONSTRAINT1},\ref{CONSTRAINT2}):
$$                                                                                                                      
\sum_{l} c_l^i f_l^{(1)}=0; \qquad \sum_{l} c_l^i \bar{f}_l^{(1)}=
\rho \chi_i
$$
$$                                                                                                                      
\sum_l c_l^i c_l^j f_l^{(0)}=\rho u_i u_j+ \rho T \delta_{ij}
$$
$$                                                                                                                      
\frac{1}{2} \sum_{l} c_l^2 f_l^{(1)}=0; \qquad \frac{1}{2} \sum_{l}
c_l^2 \bar{f}_l^{(1)}=\rho \chi_i u_i
$$
$$                                                                                                                      
\frac{1}{2}\sum_l c_l^i c_l^2 f_l^{(0)}=\left(\frac{1}{2}\rho u^2 +\frac{D}{2} \rho T \right)u_i +\rho T u_i .          
$$
With this, using the momenta of ${\cal O}(\epsilon)$ in (\ref{eq:ce}),
we can easily arrive to the following set of equations
\be\label{THEFIRSTORDER}
\begin{cases}
  \partial_{1t} \rho+ \partial_{i}(\rho u_i)=0 \vspace{0.1in}\\
  \partial_{1t}(\rho u_i)+\partial_j(\rho u_i u_j+\rho T \delta_{ij})=\frac{\chi_i}{\tau}=g_i  \vspace{0.1in}\\
  \partial_{1t}{\cal K}+\partial_{j}\left[ {\cal K} u_j+\rho T u_j
  \right] =\frac{1}{\tau}\rho \chi_i u_i= \rho g_i u_i \hspace{.2in}
\end{cases}
\ee where we have introduced the total energy of the system:
$$                                                                                                                      
{\cal K}=\left(\frac{1}{2} \rho u^2+\frac{D}{2} \rho T \right)
$$
and where we have recovered the Euler equations for a forced fluid
with the choice \be {\bm \chi}=\tau {\bm g}.  \ee The last equation
can also be written as an equation for the temperature (using the
momentum equation) in the following form \be
(\partial_{1t}+u_j \partial_j) T+\frac{1}{c_v} T (\partial_i u_i)=0;
\hspace{.2in} c_v=\frac{D}{2}.  \ee
\subsubsection{Second Order}
Using the second of (\ref{eq:ce}) and the constraints found at the
first order it is easy to derive: \be \sum_l c_l^i (\partial_{1t}
f^{(0)}_l+c_l^k \partial_k f_l^{(0)})=-\frac{1}{\tau}\sum_l c_l^i
(f^{(1)}_l- \bar{f}^{(1)}_l ) =\rho g_i.  \ee Furthermore, let us
write other useful quantities that can be derived from the explicit
expression of the expansion of the equilibrium distribution, $\bar
f_l$, and from the hydrodynamical constraints on the distribution
$f_l$ reported in (\ref{CONSTRAINT1}) and (\ref{CONSTRAINT2}):
\be\label{C2a} \sum_{l}c_l^i c_l^j c_l^k f_l^{(0)}=(\rho u_i u_j
u_k+\rho T(\delta_{ij} u_k+\delta_{ik} u_j +\delta_{jk} u_i)) \ee
\be\label{C2b} \sum_{l}c_l^i \bar{f}_l^{(2)}=0; \qquad \sum_{l}c_l^i
{f}_l^{(2)}=0 \ee \be\label{C2c} \frac{1}{2}\sum_l c_l^2 c_l^i
\bar{f}_l^{(1)}=\frac{\rho u^2 \chi_i}{2} + u_i \rho \chi_j u_j+ \rho
T \chi_i+\frac{D \rho T \chi_i}{2} \ee
\begin{eqnarray}\label{C2d}
\frac{1}{2}\sum_l c_l^2 c_l^i c_l^j \bar{f}_l^{(0)}=\frac{1}{2} \rho u_i u_j u^2+\frac{\rho T}{2} \delta_{ij} u^2 
+ \nonumber \\
2 \rho T u_i u_j +\frac{1}{2} D \rho T u_i u_j+ \left( \frac{D}{2}+1 \right) \rho T^2 \delta_{ij}
\end{eqnarray}
\be
\frac{1}{2}\sum_{l} c_l^2 \bar{f}_l^{(2)}=\frac{1}{2} \rho \chi^2+\frac{1}{2}D \rho \lambda.
\ee
We next proceed to evaluate some expressions in terms of the known results obtained at the previous order. In particular, for the momentum equation, we will have to evaluate the term:
$$                                                                              \partial_{1t} \left( \sum_l c^{i}_l c^{j}_l f_{l}^{(0)} \right)=\partial_{1t}(\rho u_i u_j+\rho T \delta_{ij}).         
$$
If we use the results obtained at order ${\cal O}(\epsilon)$ in (\ref{THEFIRSTORDER}) we obtain
\begin{eqnarray}\label{M2a}
\partial_{1t}(\rho u_i u_j+\rho T \delta_{ij})=-\partial_{k}(\rho u_i u_j u_k)-u_j \partial_i(\rho T)-  \nonumber \\
 u_i \partial_{j}(\rho T) + \rho u_j g_i+\rho u_i g_j+\delta_{ij} \rho \partial_{1t} T+\delta_{ij} T \partial_{1t} \rho\
 .
\end{eqnarray}
Next, for the momentum equation, we also have to consider
\begin{eqnarray}
\partial_{1t}\left(\sum_l c^{i}_l c^{j}_l f_{l}^{(0)} \right)+\partial_k\left( \sum_l c^{i}_l c^{j}_l c_{l}^k f_{l}^{(0\
)} \right)&=& \nonumber \\
\partial_{1t}(\rho u_i u_j+\rho T \delta_{ij})+\partial_k[\rho u_i u_j u_k + \nonumber \\
\rho T (\delta_{ij} u_k +\delta_{ik}u_j+\delta_{jk} u_i)] \nonumber
\end{eqnarray}
that can be simplified (with results of the previous order) as
\begin{eqnarray}\label{M2b}
\partial_{1t}(\rho u_i u_j+\rho T \delta_{ij}) +\hspace{.6in} \nonumber  \\
+\partial_k[\rho u_i u_j u_k +\rho T (\delta_{ij} u_k+\delta_{ik}u_j+\delta_{jk} u_i)] =\nonumber  \\
\rho T \partial_{i} u_j+ \rho T \partial_j u_i +\rho u_j g_i+\rho u_i g_j -\delta_{ij} \frac{\rho T}{c_v}(\partial_k u_k).
\end{eqnarray}
For the energy equation we will have to consider
$$                                                                                                                      
\partial_{1t} \left( \sum_l \frac{c_l^2}{2} c^{i}_l  f_{l}^{(0)} \right)=\partial_{1t}\left[\left(\frac{1}{2} \rho u^2 \
+\frac{D}{2} \rho T  \right)u_i +\rho T u_i \right]                                                                     
$$
that, again, can be evaluated using the results at previous order as
\begin{eqnarray}\label{M2c}
\partial_{1t}\left[\left(\frac{1}{2} \rho u^2 +\frac{D}{2} \rho T  \right)u_i +\rho T u_i \right]=\rho (g_k u_k) u_i+
\rho T g_i \nonumber \\
+\left(\frac{1}{2} \rho u^2 +\frac{D}{2} \rho T  \right)g_i-\partial_j  \left[u_i u_j  \left(\frac{1}{2} \rho u^2 +
\frac{D}{2} \rho T  \right)\right] \nonumber  \\
-2 \partial_k (\rho T u_i u_k)- \partial_i\left(\frac{1}{2} \rho T u^2 \right)+\rho T u_j \partial_i u_j\nonumber 
\hspace{.55in}\\
-\left(\frac{D}{2}+1 \right) T \partial_{i}(\rho T)-\frac{1}{c_v} \rho T u_i (\partial_k u_k)+\rho u_i \partial_j u_j. \
\hspace{.1in}
\end{eqnarray}
Finally, we have to consider
\begin{eqnarray}
&&\partial_{1t} \left( \sum_l \frac{c_l^2}{2} c^{i}_l f_{l}^{(0)} \right)+\partial_j \left( \sum_l 
\frac{c_l^2}{2} c^{i}_l c^{j}_l f_{l}^{(0)} \right)\nonumber \hspace{0.8in} \\
&&=\partial_{1t}\left[\left(\frac{1}{2} \rho u^2 +\frac{D}{2} \rho T  \right)u_i +\rho T u_i \right] \nonumber \\
&&+\partial_j \left[\frac{1}{2} \rho u_i u_j u^2+\frac{\rho T}{2} \delta_{ij} u^2+ 2 \rho T u_i u_j \right] \nonumber \\
\
&&+\partial_j \left[ \frac{1}{2} D \rho T u_i u_j+ \left( \frac{D}{2}+1 \right) \rho T^2 \delta_{ij} \right] \nonumber
\end{eqnarray}
that gives
\begin{eqnarray}\label{M2d}
\partial_{1t} \left( \sum_l \frac{c_l^2}{2} c^{i}_l f_{l}^{(0)} \right)+\partial_j \left( \sum_l \frac{c_l^2}{2} c^{i}_l 
c^{j}_l f_{l}^{(0)} \right) = \nonumber  \\
+\rho (g_k u_k) u_i+\rho T g_i+\left(\frac{1}{2} \rho u^2 +\frac{D}{2} \rho T  \right)g_i+\nonumber \\  
\left(\frac{D}{2}+1 \right) \rho T \partial_i T+\rho T (u_i \partial_j u_j+u_j \partial_i u_j)-\nonumber \\
-\frac{1}{c_v} \rho T u_i (\partial_k u_k).
\end{eqnarray}
We are now ready to write down the equations at this order using results in (\ref{C2a})-(\ref{C2d}) and 
(\ref{M2a})-(\ref{M2d})
\be
\begin{cases}
\partial_{2t} \rho+1/2 \partial_{i} \left(\rho g_i \Delta t \right)=0 \vspace{0.1in}\\
\partial_{2t}(\rho u_i)+\partial_{j}( \tau \rho g_i u_j+\tau \rho g_j u_i)
+\partial_{1t} \left( \frac{\rho g_i}{2} \Delta t \right)  =\nonumber \\
\left(\tau-\frac{\Delta t}{2} \right) \partial_{j} ( \rho T \partial_{i} u_j+ \rho T \partial_j u_i +\rho u_j g_i+\rho \
u_i g_j \nonumber \\
-\delta_{ij} \frac{\rho T}{c_v}(\partial_k u_k) ) \vspace{0.1in}\\
\partial_{2t}{\cal K}+\partial_{k}\left[{\cal K} \tau g_k+\tau \rho u_k g_k +\rho T g_k  \right]+\partial_{1t} 
\left( \frac{\rho g_i u_i}{2} \Delta t \right)- \nonumber \\
\left(\tau-\frac{\Delta t}{2} \right) \partial_i [ \rho (g_k u_k) u_i+\rho T g_i+{\cal K}g_i+\left(\frac{D}{2}+1 \right) 
\rho T \partial_i T\nonumber \\
+\rho T u_i \partial_j u_j+\rho T u_j \partial_i u_j-\frac{1}{c_v} \rho T u_i (\partial_k u_k) ] =  \nonumber \\
=\frac{1}{\tau} \left(\frac{1}{2} \rho \tau^2 g^2+ \frac{1}{2} D \rho \lambda \right). \hspace{.2in}
\end{cases}
\ee
Summing up all orders, we note  that we can {\it freely} add at order ${\cal O}(\epsilon^2)$ all the gradients of terms\
 ${\cal O} (g^2)$ and also double gradients of terms ${\cal O} (g)$ because they would be ${\cal O}(\epsilon^3)$. Also,\
  defining the hydrodynamic velocity as ${u}^{(H)}_i=u_i+\frac{g_i \Delta t}{2}$, we reconstruct the following equation\
s:
\be
\begin{cases}
\label{eq:final}
\partial_t \rho+\partial_{i}(\rho {u}^{(H)}_i)=0 \vspace{0.1in}\\
\partial_{t}(\rho {u}^{(H)}_i)+\partial_{j}(\rho {u}^{(H)}_i {u}^{(H)}_j)=-\partial_{i}(\rho T)+g_i\nonumber \\
+\left(\tau-\frac{\Delta t}{2} \right) \partial_j \left[\rho T \partial_i {u}^{(H)}_j+\rho T \partial_j {u}^{(H)}_i-
\delta_{ij} \frac{\rho T}{c_v} \partial_{k} {u}^{(H)}_k \right] \hspace{.1in} \vspace{0.1in}\\
\partial_{t}{\cal K}^{(H)}+\partial_{j}\left[ {\cal K}^{(H)}u^{(H)}_j+\rho T {u}^{(H)}_j \right]   =\rho g_k u_k+\nonumber \\
\frac{1}{2\tau} \left( \rho \tau^2 g^2+  D \rho \lambda \right)+ \left(\tau-\frac{\Delta t}{2} \right) \partial_i  
[ (\frac{1}{2} D+1) \rho T \partial_i T +\nonumber \\ \rho T u^{(H)}_i \partial_j u^{(H)}_j+\rho T u^{(H)}_j \partial_i u^{\
(H)}_j-\frac{1}{c_v} \rho u^{(H)}_i (\partial_k u^{(H)}_k) ] \hspace{.1in}
\end{cases}
\ee
with
$$                                                                                                                      
{\cal K}^{(H)}=\left(\frac{1}{2} \rho (u^{(H)})^2+\frac{D}{2} \rho T \right).                                           
$$
In order to recover the correct thermohydrodynamical evolution we need to obtain the correct forcing in the equation for the total energy in terms of the hydrodynamical velocity fields, i.e.
$$                                                                                                                      
\rho g_k u_k+\frac{1}{2\tau} \left( \rho \tau^2 g^2+  D \rho \lambda \right)=\rho g_k u^{(H)}_k=\rho g_k \left( 
u_k+\frac{\Delta t g_k}{2}\right)                                                                                              
$$
that leads to
\be
\lambda=\frac{\tau(\Delta t-\tau)g^2}{D}.
\ee
In conclusions, expressing everything in terms of the hydrodynamical fields, it is easy to realize that the final expression (\ref{eq:final}) coincides with the one given in the body of the article (\ref{eq:FNS1}). Notice that up to now we have used a single-time relaxation LBM, as given by (\ref{MAIN}). Therefore, the final Fourier-Navier-Stokes equations are constrained to describe fluids with unit Prandtl numbers, $Pr = \nu/(k/c_p) =1 $. It is possible to generalize the
approach by using a multi-relaxation time version of the same algorithm \cite{Shan07}. Even though, in the latter case, there exists a small mismatch in the viscous dissipation term appearing in the energy balance.

\section{APPENDIX B}

In this appendix we detail the technical steps leading to the desired
hydrodynamical boundary conditions for the physical systems analyzed
in the paper, i.e. an ideal gas under the effect of gravity ${\bm
  g}=(0,-g)$ acting along the negative $z$ direction (i.e. $g$ is
positive). Similar ideas can be applied to the case of a generic
volume or internal force acting also in the stream-wise $x$
direction. For the sake of concreteness we explicitly report the case
of the lower boundary condition with the upper boundary condition
being a straightforward generalization. Let us call the post streaming populations $f^{*}_l$ while keeping $f^{(*,pre)}_l$ to identify
the pre streaming populations. Moreover, all the populations will
also undergo collisions and therefore there will be a net gain of
momentum so that the hydrodynamic fields will be the average of pre
and post collisions. For a given computational boundary, there are 3
layers of points labeled by ${\bm x}^*$ from now on (see also figure
 \ref{fig:WALLD2Q37}), where some unknown populations have to be set soon after the streaming step. We use the freedom to set these
populations in such a way that the measured hydrodynamic quantities
such as the stream-wise ($u_x^{(H)}$) and vertical ($u_z^{(H)}$)
velocity and also the temperature ($T^{(H)}$) are fixed to some given
boundary conditions on those lattice layers. The conditions to be
fulfilled up to the second order in the Chapman-Enskog expansion
(see also previous appendix) are 
\be {u}_x^{(H)}({\bm x}^*)=\frac{1}{
  \rho({\bm x}^*)}\sum_l f^{*}_l({\bm x}^*) {c}^x_l \ee \be
{u}_z^{(H)}({\bm x}^*)=\frac{1}{ \rho({\bm x}^*)}\sum_l f^{*}_l({\bm
  x}^*) {c}^z_l -\frac{\Delta t}{2} {g} \ee
\begin{eqnarray}
\frac{D}{2} T^{(H)} ({\bm x}^* )+\frac{1}{2}(({u}^{(H)})^2+({v}^{(H)})^2)({\bm x}^* )= \nonumber  \\
= \frac{1}{ 2 \rho({\bm x}^*) }\sum_l  f^{*}_l({\bm x}^*) c_l^2.
\end{eqnarray}
In the following we  show how to determine the unknown populations on the first three layers (those coming --after streaming-- from node outside the domain) in order to set the vertical velocity to zero on layer 3, with any temperature and stream-wise velocities:
$$                                                                                                                      
\begin{cases}                                                                                                           
u_z^{(H)}(z=3) =0 \\                                                                                                    
u_x^{(H)}(z=3) = u_3\\                                                                                                  
T^{(H)}(z=3) = T_3.                                                                                                     
\end{cases}                                                                                                             
$$
Similarly we can fix any desired profile for temperature and velocity on layers $1,2$:
$$
\begin{cases}                                                                                                           
u_z^{(H)}(z=2) = v_2;\qquad u_z^{(H)}(z=1) =v_1\\                                                                       
u_x^{(H)}(z=2) = u_2;\qquad u_x^{(H)}(z=1) =u_1\\                                                                       
T^{(H)}(z=2) = T_2;\qquad T^{(H)}(z=1) =T_1.                                                                            
\end{cases}                                                                                                             
$$
We will define only the case of homogeneous boundary conditions along
the stream-wise component but the method is general and can deal
also non-homogeneous cases.  Imposing a given set of boundary
conditions means defining the set of {\it unknown} outgoing
populations in the first three layers in terms of the set of in-going and outgoing {\it known} populations such that mass is conserved and the hydrodynamical fields defined above are the wanted ones. \\
In this way, if the computational boundary extends from the mesh point
$z=1$ up to $z=L_z$, the real physical domain is between mesh points
$z=3$ and $z=L_z-2$, i.e. it is in these points that we exactly verify
the condition of no-slip, no normal velocity and given temperature
for the hydrodynamical fields on the solid walls. Fields at points
$z=1,2$ and $z=L_z-1,L_z-2$ may be used to better stabilize the
algorithm close to the boundaries. All details refer to the 37 speed
model $D2Q37$.

\subsubsection*{Layer 1}

As evident from figure \ref{fig:WALLD2Q37}  we have to determine some 'outer' post streaming populations ($l=2,10,18...$) whereas other post streaming populations ($l=4,12,20...$) are known. To keep a compact notation, let us also introduce the subsets $I^{(1)},U^{(1)}$ and $I^{(1)}_{0}$ which are identified by the following conditions
$$I^{(1)} =\{{\bm c}_l, c_{l}^{z} < 0  \}; \hspace{.2in} U^{(1)} =\{{\bm c}_l, c_{l}^{z} > 0  \}$$
$$                                                                                                                      
I^{(1)}_0 =\{{\bm c}_l, c_{l}^z \le 0  \}.$$
We choose to define the 'outer' populations in the layer $1$ as
\be\label{OUTER}
f^{(1,*)}_{l}= \frac{N}{\sum_{l \in U^{(1)}} \phi^{(1)}_l}  \phi^{(1)}_l  \hspace{.2in} l \in U^{(1)}
\ee
with $N$ a constant and $\phi^{(1)}_l$ a suitable population that we choose in the form
\be\label{OUTER2}
\phi^{(1)}_l=1+{\bm c}_l \cdot {\bm p}^{(1)}+\frac{1}{2} c_l^2 {E}^{(1)}
\ee
where ${p}_x^{(1)}$,${p}_z^{(1)}$ and ${E}^{(1)}$ are unknown at this level and must be chosen in such a  way that the hydrodynamical temperature and momentum {\it exactly} reproduce the desired values on this layer, $T_1,u_1,v_1$. Also, mass conservation should be fulfilled. This latter condition is naturally imposed by setting
$$                                                                                                                      
N=\sum_{l \in I^{(1)}} f^{(1,*,pre)}_l.                                                                                 
$$
The requirement that $T_1,u_1,v_1$ are exactly reproduced leads to the following system of equations
\be\label{SYSTE}
\begin{cases}
{u}_1=\frac{1}{M_p}\sum_l  f_l^{*} {c}_{l}^x  \\
{v}_1=\frac{1}{M_p}\sum_l  f_l^{*} {c}_{l}^z  -\frac{\Delta t}{2} g\\
T_1= \frac{1}{M_p D}\sum_l  f_l^{*} c_l^2 +\frac{1}{D}(u_1^2+v_1^2  )
\end{cases}
\ee
where we have defined the post streaming mass as
$$                                                                                                                      
M_p=N+\sum_{l \in I^{(1)}_0} f^{(1,*)}_l.                                                                               
$$
In the $\sum_l  f_l^*$ of system (\ref{SYSTE}) we have known populations coming from the bulk but also 'outer' populations\
 to be determined with (\ref{OUTER}) and (\ref{OUTER2}). The resulting system is therefore an algebraic system for  ${p}_x^{(1)}$,${p}_z^{(1)}$ and ${E}^{(1)}$. We have solved the system whose final solution is
$$                                                                                                                      
p^{(1)}_z=\frac{-c_3 d_2+c_2 d_3}{-a_3 c_2+a_2 c_3}                                                                     
$$
$$                                                                                                                      
p^{(1)}_x=\frac{a_2 c_3 d_1-a_2 c_1 d_3-a_3 c_2 d_1-c_3 a_1 d_2+c_1 a_3 d_2+a_1 c_2 d_3}{b_1 (a_3 c_2-a_2 c_3)}         
$$
$$                                                                                                                      
E^{(1)}=\frac{-a_3 d_2+a_2 d_3}{a_3 c_2-a_2 c_3}                                                                        
$$
where
$$                                                                                                                      
a_1=26(\tilde{p}_x-O_x)r; \hspace{.2in} b_1=-40 N r^2                                                                   
$$
$$                                                                                                                      
c_1=47 (\tilde{p}_x-O_x) r^2; \hspace{.2in} d_1=15 (\tilde{p}_x-O_x)                                                    
$$
$$                                                                                                                      
a_2=26 (\tilde{p}_z-O_z)r-54 N r^2; \hspace{.2in} c_2=47 (\tilde{p}_z-O_z)r^2-91 N r^3                                  
$$
$$
d_2=15 (\tilde{p}_z-O_z)-26 N r; \hspace{.2in} a_3=26 (\tilde{E}-O_e)r-91 N r^3 \hspace{.2in}
$$                                                                                                                      
$$
c_3=47 (\tilde{E}-O_e) r^2-\frac{367}{2} N r^4;  \hspace{.2in}  d_3=15 (\tilde{E}-O_e)-47 N r^2
$$
with
$$
\tilde{p}^{(1)}_x=M_p {u}_1; \hspace{.2in} \tilde{p}^{(1)}_z=M_p {v}_1+ \frac{1}{2} M_p g \Delta t
$$                                                                                                                      
$$
\tilde{E}={T}_1 M_p+\frac{1}{2 M_p}((\tilde{p}^{(1)}_x)^2+(\tilde{p}^{(1)}_z)^2)
$$                                                                                                                      
and                                                                                                                     
$$
O_x=\sum_{l \in I^{(1)}_0} c_{l}^x f^{(1,*)}_l; \hspace{.2in} O_z=\sum_{l \in  I^{(1)}_0} c_{l}^z f^{(1,*)}_l
$$                                                                                                                      
$$
O_e=\sum_{l \in  I^{(1)}_0} \frac{1}{2} c_{l}^2f^{(1,*)}_l.
$$
In the above $r$ is the lattice constant whose value for the $D2Q37$
model is $r \sim  1.1969$ 
\cite{Shan07}.

\subsubsection*{Layer 2}
Situation goes similarly with respect to the previous layer.  We new have to define the subsets $I^{(2)},U^{(2)}$ and $I^{(2)}_0$ as
$$I^{(2)} =\{{\bm c}_l, c_{l}^z < -r  \}; \hspace{.2in} U^{(2)} =\{{\bm c}_l, c_{l}^z > r  \}$$
$$I^{(2)}_0 =\{{\bm c}_l, c_{l}^z \le r  \}.$$
We then identify some coarse grained quantities as
$$                                                                                                                      
N=\sum_{l \in I^{(2)}} f^{(2,*)}_l; \hspace{.2in} M_p=N+\sum_{l \in I^{(2)}_0} f^{(2,*)}_l                              
$$
ad define some local momentum and energy fields
$$
\tilde{p}^{(2)}_x=M_p {u}_2
$$                                                                                                                      
$$
\tilde{p}^{(2)}_z=M_p {v}_2+ \frac{1}{2} M_p g \Delta t
$$                                                                                                                      
$$
\tilde{E}^{(2)}={T}_2 M_p+\frac{1}{2 M_p}((\tilde{p}_x^{(2)})^2+(\tilde{p}_z^{(2)})^2).
$$                                                                                                                      
We next define                                                                                                          
$$
O_x=\sum_{l \in I^{(2)}_0} c_{l}^x f^{(2,*)}_l; \hspace{.2in} O_z=\sum_{l \in  I^{(2)}_0} c_{l}^z f^{(2,*)}_l
$$                                                                                                                      
$$
O_e=\sum_{l \in  I^{(2)}_0} \frac{1}{2}  c_{l}^2f^{(2,*)}_l
$$                                                                                                                      
$$a_1=19 (\tilde{p}_x-O_x) r; \hspace{.2in} b_1=-12  N r^2 $$                                                        
$$c_1= \frac{59}{2}(\tilde{p}_x-O_x) r^2; \hspace{.2in}  d1=8 (\tilde{p}_x-O_x) $$
$$ a_2=19 (\tilde{p}_z-O_z) r-47 N r^2; \hspace{.2in} c_2= \frac{59}{2}(\tilde{p}_z-O_z)r^2- \frac{147}{2} N r^3 $$
$$d_2=8 (\tilde{p}_z-O_z)-19 N r; \hspace{.2in} a_3=19 (\tilde{E}-O_e) r- \frac{147}{2} N r^3 \hspace{.2in}$$
$$c_3= \frac{59}{2} (\tilde{E}-O_e) r^2-\frac{475}{4}N r^4;  \hspace{.2in} d_3=8 (\tilde{E}-O_e)-\frac{59}{2} N r^2. $$
In terms of these constants and parameters we can set
$$                                                                              p^{(2)}_z=\frac{-c_3 d_2+c_2 d_3}{-a_3 c_2+a_2 c_3}                              $$
$$
p^{(2)}_x=\frac{a_2 c_3 d_1-a_2 c_1 d_3-a_3 c_2 d_1-c_3 a_1 d_2+c_1 a_3 d_2+a_1 c_2 d_3}{b_1 (a_3 c_2-a_2 c_3)}
$$                                                                                                                      
$$
E^{(2)}=\frac{-a_3 d_2+a_2 d_3}{a_3 c_2-a_2 c_3},
$$
construct suitable populations
$$
\phi^{(2)}_l=1+{\bm c}_l \cdot {\bm p}^{(2)}+\frac{1}{2} c_l^2 E^{(2)}
$$
and define the outer populations in the layer $2$ as
$$                                                                                                                      
f^{(2,*)}_{l}= \frac{N}{\sum_{l \in U^{(2)}} \phi^{(2)}_l}
\phi^{(2)}_l \hspace{.2in} l \in U^{(2)}
$$
that is enough to set the hydrodynamic velocity to $u_2$ and $v_2$
while keeping the hydrodynamic temperature to $T_2$.

\subsubsection*{Layer 3}
\begin{figure}
\begin{center}
\includegraphics[scale=0.5]{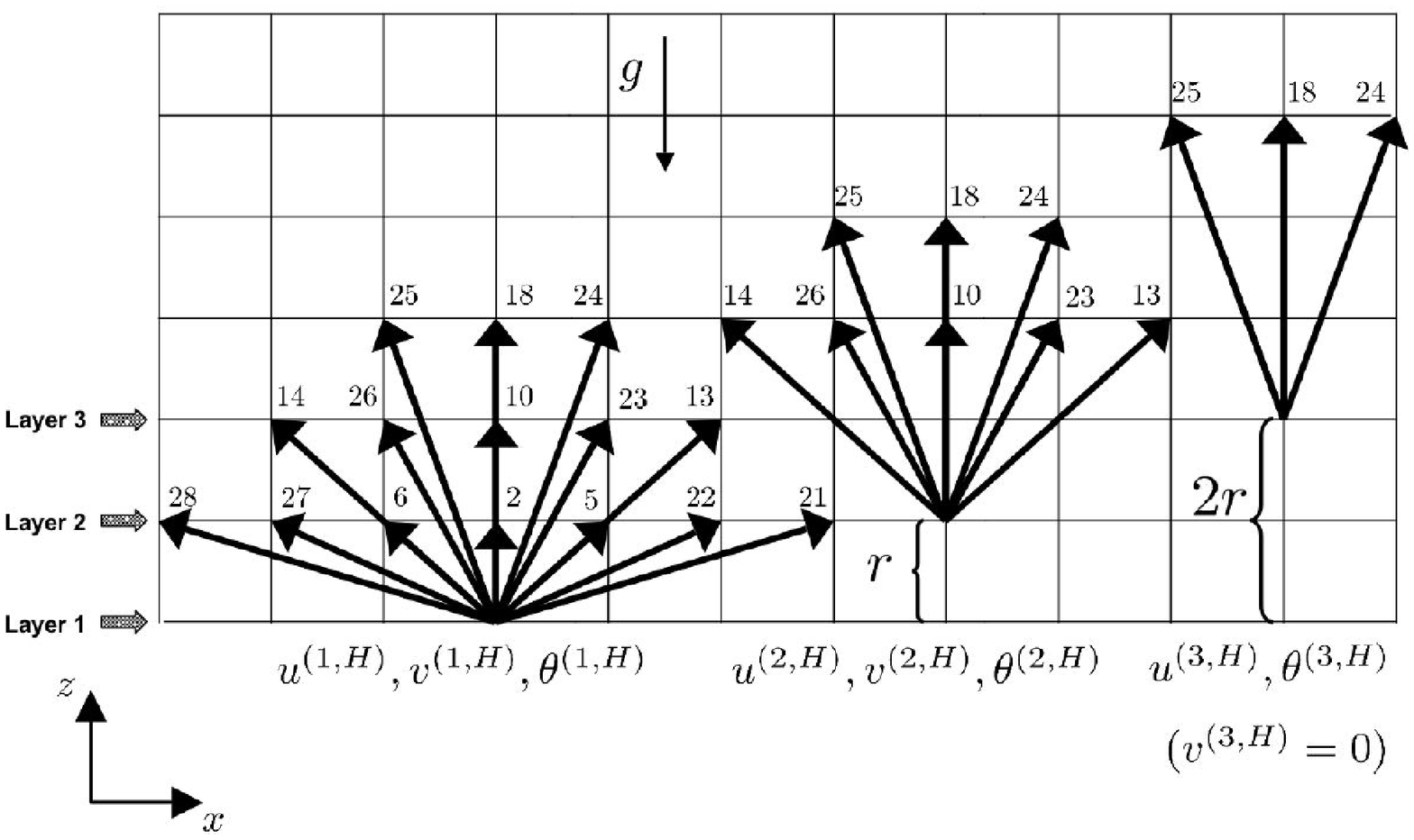}
\caption{Scheme for the lower boundary layer for the simulation of
  thermal flows under the effect of gravity. The relationship
  between real and velocity lattice is set by $\Delta x= r \Delta t$
  with $\Delta x$ and $\Delta t$ space and time discretizations, and
  $r$ the lattice constant whose value is $r \sim
  1.1969$. the locations at $r$ and $2r$
  indicated in this figure correspond to the locations $z=2$ and $z=3$
  discussed in the text.}
\label{fig:WALLD2Q37}
\end{center}
\end{figure}

As also evident from the figure \ref{fig:WALLD2Q37}, only 3
populations are unknown on the third layer (they are populations
$l=24$,$25$,$18$). In this way we do not have enough freedom to choose
the desired hydrodynamic velocities and temperature. It is anyhow
possible to require a zero vertical hydrodynamic velocity ($v_3=0$)
with a generic stream-wise hydrodynamic velocity and temperature
($u_3, T_3$). Again, let us introduce the following sets
$$U^{(3)} =\{{\bm c}_l, c_{l}^z > 2r  \}; \hspace{.2in} I^{(3)}_0 =\{{\bm c}_l, c_{l}^z \le 2r  \}.$$
The boundary condition for the unknown populations is set as
$$                                                                                                                      
f^{(3,*)}_{l}= \frac{N}{\sum_{l \in U^{(3)}} \phi^{(3)}_l}  \phi^{(3)}_l  \hspace{.2in} l \in U^{(3)}                   
$$
$$                                                                              \phi^{(3)}_l=1+{c}_{l}^x  {p}^{(3)}_x+\frac{1}{2} c_l^2 E^{(3)}                   $$
and we choose ${p}^{(3)}_x$ and $E^{(3)}$ to set the desired hydrodynamical stream-wise velocity ($u_3$) and temperatur\
e ($T_3$) while keeping the vertical hydrodynamical velocity to zero. The resulting algebraic system is solved with the\
 solution
$$                                                                              E^{(3)}=-\frac{d2}{b2}; \hspace{.2in} p^{(3)}_x=\frac{b_1 d_2-d_1 b_2}{a_1 b_2}  $$
with
$$                                                                              a_1=-2 N r^2; \hspace{.2in} b_1=\frac{29}{2}(\tilde{p}_x-O_x) r^2               $$
$$                                                                              d_1=3 (\tilde{p}_x-O_x); \hspace{.2in} b_2=\frac{29}{2}(\tilde{E}-O_e) r^2-\frac{281}{4} N r^4                          
$$
$$                                                                              d_2=3 (\tilde{E}-O_e)-\frac{29}{2} N r^2                                        $$
where
$$                                                                              O_x=\sum_{l \in I^{(3)}_0} c_{l}^xf^{(3,*)}_l;  \hspace{.2in} O_e=\sum_{l \in  I^{(3)}_0} \frac{1}{2}  c_{l}^2f^{(3,*)}_l                                                                              $$
$$                                                                              \tilde{p}^{(3)}_x=M_p {u}_3                                                     $$
$$
 \tilde{E}^{(3)}={T}_3 M_p+\frac{1}{2 M_p}((\tilde{p}_x^{(3)})^2+(\tilde{p}_z^{(3)})^2)
$$                                                                              $$
\tilde{p}^{(3)}_z=3 N r+\sum_{l \in I^{(0)}_3} c_{l}^z f_l^{(3,*)}
$$                                                  
$$
M_p=N+\sum_{l \in I^{(3)}_0} f^{(3,*)}_l.
$$                                                                  
$$
N=-\frac{A_1}{3 r}+\frac{1}{2} \frac{A_2}{3 r}  g  \Delta t
$$
\be\label{APPEAR}
A_1=\sum_{l \in I^{(0)}_3} c_{l}^z f_l^{(3,*)}; \hspace{.2in} A_2=\sum_{l \in I^{(0)}_3}  f_l^{(3,*)} - f_0^{(3,*)}.
\ee
This whole algorithm for layer 3 now is ensuring a  zero vertical hydrodynamical velocity and arbitrary $u_3$ and $T_3$. Still, mass conservation is not fulfilled and to do that we need to redefine the rest population appearing in 
(\ref{APPEAR}) as
$$                                                                              f_0^{(3,*)}=f_0^{(3,*,pre)}-N+\sum_{l, c_{l}^z=-3r} f_l^{(3,*)}.                      $$

\clearpage

\clearpage

\begin{table*}
\begin{center}
\begin{tabular}{|c | c c c c c c c c c c |}
  \hline & $At$ & $L_x$ & $L_z$ & $L_{ad}$& $\nu$ & $g$ & $T_u$ & $T_d$ & $N_{conf}$ & $\tilde{\tau}$ \\
\hline run A & $0.05$ & 800 & 2400 & $4 \times 10^3$ & 0.001 & $5 \times 10^{-5}$ & $0.95$ & $1.05$ & 50 & $1.8 \times 10^4$ \\
run B &$0.4$& 1664 & 4400 &$1.6 \times 10^4$ &0.1 & $1 \times 10^{-4} $ &
$0.6$ & $1.4$ & 35 & $ 6.5 \times 10^3$ \\ \hline
\end{tabular}
\caption{Parameters for the two sets of \RT\ run. Atwood number, $At=(T_d-T_u)/(T_d+T_u)$; 
Adiabatic Length, $L_{ad} = (T_d-T_u)c_p/g$ ($c_p=2$); viscosity $\nu$; gravity $g$; 
temperature in the upper half region, $T_u$; temperature in the lower half region, $T_d$; 
number of separate \RT\ run $N_{conf}$; normalization time,
$\tilde{\tau}=\sqrt{L_x/(g\;At)}$ (not to be confused with the
relaxation time of the lattice Boltzmann model (\ref{MASTERSHIFT3})). Given the parameters here used, the typical resolution obtained is
good enough to get an agreement better than a few per cent on the
global exact balance between kinetic energy growth and the sum of 
dissipation plus buyoancy force.
\label{table:param} }
\end{center}
\end{table*}

\clearpage

\maketitle

\begin{thebibliography}{81}

\bibitem{SC1}  X. Shan X. \& H. Chen. Lattice Boltzmann model for simulating flows with multiple phases and components. {\it Phys. Rev E} {\bf 47}, 1815 (1993)

\bibitem{Yeo} M.R. Swift, W.R.  Osborn \&  J.M. Yeomans.  Lattice Boltzmann Simulation of Nonideal Fluids. {\it Phys. Rev. Lett.} {\bf 75}, 830 (1995)

\bibitem{Wagb}Q. Li  \& A.J. Wagner. Symmetric free-energy-based multicomponent lattice Boltzmann method. {\it Phys. Rev. E} {\bf 76}, 036701 (2007)

\bibitem{HeLuo}X. He X. \& L.S. Luo. Theory of the Lattice Boltzmann Method: from the Boltzmann equation to the Lattice Boltzmann equation. {\it Phys. Rev. E} {\bf 56}, 6811 (1997)

\bibitem{Ladd} A.J.C. Ladd. Numerical simulations of particulate suspensions via a discretized Boltzmann equation. 2. Numerical Results. {\it J. Fluid. Mech.}{\bf 271}, 311 (1994)



\bibitem{Hartingb} J. Harting, C. Kunert \& J. Hyvaluoma. Lattice Boltzmann simulations in microfluidics: probing the no-slip boundary condition in hydrophobic, rough, and surface nanobubble laden microchannels. {\it  Microfluidics and Nanofluidics} {\bf 8}, 1 (2009)

\bibitem{ladd-review} B. Duenweg \& A. J. C. Ladd. Lattice Boltzmann simulations of soft matter systems. 
{\it Adv. Poly. Sci.} {\bf 221}, 89-166  (2009) 

\bibitem{jos} R. R. Nourgaliev, T. N. Dinh, T. G. Theofanous \& D. Joseph. The Lattice Boltzmann equation method: theoretical interpretation, numerics and implications. {\it Int. J. Multiphase Flow} {\bf 29}, 117 (2003) 
  
\bibitem{Gladrow} D. Wolf-Gladrow. {\it Lattice-Gas Cellular Automata And Lattice Boltzmann Models}. Springer, New York (2000)

\bibitem{BSV}R. Benzi, S. Succi  \& M. Vergassola. The lattice Boltzmann equation: theory and applications. {\it Phys. Rep.} {\bf 222}, 145 (1992)

\bibitem{Chen}S. Chen \& G. Doolen. Lattice Boltzmann method for fluid flows. {\it Annu. Rev. Fluid Mech.} {\bf 30}, 329 (1998)

\bibitem{BGK54}P.-L. Bathnagar, E. Gross  \& M. Krook. A model for collision processes in gases. {\it Physical review} {\bf 94}, 511 (1954)

\bibitem{ShanHe98}X. Shan \& X. He. Discretization of the Velocity Space in the Solution of the Boltzmann Equation. {\it Phys. Rev. Lett.} {\bf 80}, 65 (1998)

\bibitem{Martys98}N.S. Martys, X. Shan  \& H. Chen. Evaluation of the extrenal force term in the discrete Boltzmann equation. {\it Phys. Rev. E} {\bf 58}, 6865 (1998) 

\bibitem{Shan06} X. Shan, F. Yuan \& H. Chen. Kinetic theory representation of hydrodynamics: a way beyond the Navier­Stokes equation. {\it Jour. Fluid Mech.} {\bf 550}, 413 (2006)

\bibitem{HeDoolen01} X. He  \& G. Doolen. Thermodynamic Foundations of Kinetic Theory and Lattice Boltzmann Models for Multiphase Flows. {\it Jour. Stat. Physics} {\bf 107}, 309 (2001)

\bibitem{pre.nostro} R. Benzi, L. Biferale, M. Sbragaglia, S. Succi \&  F. Toschi. Mesoscopic Modelling of a Two-Phase Flow in Presence of the Boundaries: the Contact Angle.  {\it Phys. Rev. E} {\bf 74}, 021509 (2006)

\bibitem{prl.nostro} M. Sbragaglia, R. Benzi, L. Biferale, S. Succi \& F. Toschi.  Surface roughness-hydrophobicity coupling in microchannel and nanochannel flows.  {\it Phys. Rev. Lett. } {\bf 97},  204503  (2006)

\bibitem{Harting}  J. Hyvaluoma \& J. Harting. Slip flow over structured surfaces with entrapped microbubbles. {\it   Phys. Rev. Lett.} {\bf 100}, 246001 (2008)

\bibitem{LL} P. Lallemand \& L. S. Luo. Theory of the lattice Boltzmann method: Acoustic and thermal properties in two and three dimensions. {\it Phys. Rev. E}
 {\bf 68}, 036706 (2003) 

\bibitem{Karlin} N.I. Prasianakis, I.V. Karlin, J. Mantzaras \& K.B.Boulouchos.
 Lattice Boltzmann method with restored Galilean invariance.  {\it Phys. Rev. E} {\bf 79}, 066702 (2009)

\bibitem{Prasianakis}  N.I. Prasianakis \& I.V. Karlin. Lattice Boltzmann method for thermal flow simulation on standard lattices.  {\it Phys. Rev. E} {\bf 76}, 016702 (2007)


\bibitem{Sofonea}V.  Sofonea. Implementation of diffuse reflection boundary conditions in a thermal lattice Boltzmann model with flux limiters.  {\it Jour. Comp. Phys.} {\bf 228}, 6107  (2009)

\bibitem{Gonnella} G. Gonnella A. Lamura \& V.Sofonea. Lattice Boltzmann simulation of thermal nonideal fluids. {\it Phys. Rev. E} {\bf 76}, 036703 (2007)

\bibitem{watari} M. Watari. Velocity slip and temperature jump simulations by the three-dimensional thermal finite-difference lattice Boltzmann method.  
{\it Phys. Rev. E } {\bf 79}, 066706 (2009)

\bibitem{Philippi06} P.C. Philippi et al. From the continuous to the lattice Boltzmann equation: The discretization problem  and thermal models. {\it Phys. Rev. E} {\bf 73}, 056702 (2006)

\bibitem{Nie08} X. Nie, X. Shan \& H. Chen. Thermal lattice Boltzmann model for gases with internal degrees of freedom. {\it Phys. Rev. E} {\bf 77}, 035701(R) (2008)

\bibitem{Meng09} J. Meng and Y. Zhang. Accuracy Analysis of high-order
  lattice Boltzmann models for rarified gas flows. {\it arXiv:0908.4520v2} (2009)

\bibitem{Shan07} X. Shan \& H. Chen. A general multi-relaxation-time Boltzmann collision model. {\it Int. Jour. of Modern Physics C} {\bf 18}, 635-643 (2007)

\bibitem{Surmas09}  R. Surmas, C.E. Pico Ortiz \& P.C. Philippi. Simulating thermohydrodynamics by finite difference solutions of the Boltzmann equations.  {\it Eur. Phys. J. Special Topics} {\bf 171}, 81-90 (2009)

\bibitem{boundary}  S. Ansumali \& I. Karlin.  Kinetic boundary conditions in the lattice Boltzmann method. {\it Phys. Rev. E} {\bf 66}, 026311  (2002)

\bibitem{Siebert08} D. N. Siebert, L. A. Hegele, \& P. C. Philippi. Lattice Boltzmann equation linear stability analysis: Thermal and athermal models. {\it Phys. Rev. E} {\bf 77}, 026707 (2008)


\bibitem{JFM.nostro} M. Sbragaglia et al. Lattice Boltzmann method with self-consistent thermo-hydrodynamic equilibria.  {\it J. Fluid Mech.} {\bf 628}, 299 (2009)

\bibitem{spiegel1} E.A. Spiegel. Convective instability in a comprssible atmosphere. {\it Astrophys. J.} {\bf 141}, 1068  (1965)


\bibitem{spiegel3} E.A. Spiegel \& G. Veronis.  On the Boussinesq approximation for a compressible fluid. {\it Astrophys. J.}  {\bf 131}, 442  (1960)

\bibitem{Lohse} G. Ahlers, S. Grossmann \& D. Lohse. Heat transfer and large-scale dynamics in turbulent Rayleigh-Benard convection.  {\it Rev Mod. Phys.} {\bf 81}, 503-537 (2009)


\bibitem{Buick00}J.M. Buick \& C.A. Greated. Gravity in a lattice Boltzmann model. {\it Phys. Rev E} {\bf 61}, 5307 (2000)

\bibitem{Guo02}Z. Guo, C. Zheng \& B. Shi.  Discrete lattice effects on the forcing term in the lattice Boltzmann method, {\it Phys. Rev. E} {\bf 65}, 046308 (2002)

\bibitem{mazzino} A. Celani, A. Mazzino, P. Muratore-Ginanneschi and
  L. Vozella. Phase-field model for the Rayleigh--Taylor instability
  of immiscible fluids. {\it J. Fluid Mech} {\bf 622}, 115-134 (2009)
 
\bibitem{Siebert07} D.N. Siebert, L.A. Hegele, R. Surmas, L.O. Emerich Dos Santos\& P.C. Philippi. Thermal Lattice Boltzmann in two dimensions. {\it Int. Jour. of Modern Physics C} {\bf 18}, 546 (2007)


\bibitem{gauthier} J. Frolich \& S. Gauthier. Numerical investigations from compressible to isobaric Rayleigh-Benard convection in two dimensions. {\it Eur. J. Mech. B/Fluids} {\bf 12} 141 (1993) 

\bibitem{robinson} F. Robinson \& K. Chan. Non-Boussinesq simulations of Rayleigh-Benard convection in a perfect gas. {\it Phys. Fluids} {\bf 16} 1321 (2004)

\bibitem{note} Notice that this can be exactly implemented in the LBM by redefining the  relaxation  time at each iteration such as  $(\tau-\Delta t)/2. = const. /(\T^{(H)}\rho)$

\bibitem{landau} L.D. Landau \& E.M. Lifshitz. {\it Fluid Mechanics}. Pergamon Press.

\bibitem{conv-astro} J. Thomas, N. Weiss \& S. Tobias. Downward pumping of magnetic flux as the cause of filamentary structures in sunspot penumbrae. {\it Nature} {\bf 420} 390  (2002)

\bibitem{conv-astro1} N.H. Brummell.  Turbulent compressible convection with rotation.{\it  Proceedings of the International Astronomical Union} {\bf 2},  417 (2006)

\bibitem{conv-lab1} G. Ahlers, F. F. Araujo, D. Funfschilling, S. Grossmann \& D. Lohse.  Non-Oberbeck-Boussinesq Effects in Gaseous Rayleigh-Benard Convection. {\it Phys. Rev. Letters } {\bf 98}, 054501 (2007)

\bibitem{conv-lab2} G. Ahlers, B. Dressel, J. Oh ad W. Pesch. Strong non-Boussinesq effcts near  the onset of convection in a fluid near its critical point. {\it J. Fluid. Mech.}, to appear (2009)

\bibitem{conv-lab3}J.Zhang, X.L. Wu \& K-Q Xia. Density fluctuations in strongly stratified two-dimensional turbulence. {\it Phys. Rev. Lett.} {\bf 94} 174503.
(2005)

\bibitem{chandra} S. Chandrasekhar. {\it Hydrodynamic and Hydromagnetic Stability}. Oxford Clarendon Press. 

\bibitem{gough} D.O. Gough, D.R Moore, E.A. Spiegel \& N.O. Weiss. Convective instability in a comprssible atmosphere. II. {\it Astrophys. J.} {\bf 206}, 536 (1976)

\bibitem{graham} E. Graham. Numerical simulation of two-dimensional compressible convection. {\it J. Fluid Mech.} {\bf 70}, 689 (1975)

\bibitem{verzicco} R.J.A.M. Stevens, R. Verzicco \& D. Lohse. Radial boundary layer structure and Nusselt number in  Rayleigh-Benard convection. {\it arXiv:0905.0379v1}. 

\bibitem{CHEN} H. Chen, C. Teixeira \& K. Molvig. Realization of fluid boundary
conditions via discrete Boltzmann dynamics. {\it Int. J. Mod. Phys. C} {\bf 9} 1281-1292 (1998)

\bibitem{kazu1} K. Sugiyama, E. Calzavarini, S. Grossmann \& D. Lohse. Non-Oberbeck-Boussinesq effects in two-dimensional Rayleigh-Bènard convection in glycerol. {\it Europhys. Lett. 80}, 34002 (2007)

\bibitem{kazu2} K. Sugiyama, E. Calzavarini, S. Grossmann \& D. Lohse. Flow organization in two-dimensional non-OberbeckBoussinesq Rayleigh-B\'enard convection in water.  {\it J. Fluid Mech.}  {\bf 637}, 105-135 (2009)

\bibitem{rt2} J.D. Lindl. Inertial Confinement Fusion (Springer-Verlag, New-York) 1998.

\bibitem{rt3}   M. Zingale, S.E. Woosley, C.A. Rendleman, M.S. Day  \& J.B. Bell. Three-dimensional Numerical Simulations of Rayleigh-Taylor Unstable Flames in Type Ia Supernovae.  {\it Astrophys. J.}  {\bf 632},  1021 (2005)
. 
\bibitem{rt.review} D.H. Sharp. An overview of Rayleigh-Taylor instability.  {\it Physica D} {\bf 12}, 3 (1084)

\bibitem{rt4} G. Dimonte et al. A comparative study  of the Rayleigh-Taylor instability using high-resolution three-dimensional  numerical simulations: The Alpha group collaboration. {\it Phys. Fluids} {\bf 16}   1668 (2004) 

\bibitem{jot} D. Livescu et al. High Reynolds numbers Rayleigh-Taylor turbulence. {\it J. Turbul.} {\bf 10} num. 13, 1 (2009)

\bibitem{boffi} G. Boffetta, A. Mazzino, S. Musacchio \& L. Vozzella. Kolmogorov scaling and intermittency in Rayleigh-Taylor turbulence. {\it Phys. Rev E}
{\bf 79} 065301 (2009) 

 \bibitem{chertkov} M. Chertkov. Phenomenology of Rayleigh-Taylor Turbulence. {\it Phys. Rev. Lett.} {\bf 91} 115001. (2003)

\bibitem{cabot} W. Cabot. Comparison of two- and three-dimensional simulations of miscible Rayleigh-Taylor instability. {\it Phys Fluids} {\bf 18}, 045101 (2006)

\bibitem{bernstein} I.B. Bernstein \& D. L. Book, Effect of compressibility on the Rayleigh-Taylor instability. {\it Phys Fluids} {\bf 26}, 453 (1983) 


\bibitem{gauthier3}   M.-A. Lafay, B. Le Creurer \& S. Gauthier. Compressibility effects on the Rayleigh-Taylor instability between miscible fluids. {\it Europhys. Lett.}  {\bf 79}, 64002 (2007)

\bibitem{gauthier.priv} S. Gauthier, private communication. 

\bibitem{gauthier2} B. Le Creurer \& S. Gauthier. A return toward equilibrium in a 2d Rayleigh-Taylor instability for compressible fluids with a multidomain adaptive Chebyshev method.  {\it Theor Comput. Fluid Dyn.}   {\bf 22}, 125 (2008)


\bibitem{celani1} A. Celani, A. Mazzino \& L. Vozella. Rayleigh-Taylor turbulence in two dimensions.  {\it Phys. Rev. Lett.} {\bf 96}, 134504. (2006)

\bibitem{celani2} A. Celani, T. Matsumoto, A. Mazzino \& M. Vergassola. Scaling and universality in turbulent convection. {\it Phys. Rev. Lett.} {\bf 88}, 054503 (2002)


\bibitem{cook} J.R. Ristorcelli \& T.T. Clark.  Rayleigh-Taylor turbulence: self-similar analysis and direct numerical simulations. {\it J. Fluid Mech.}
 {\bf 507}, 213  (2004)

 
\bibitem{cook2} W.H. Cabot \& A. W. Cook. Reynolds number effects on Rayleigh-Taylor instability with possible implications for type-Ia supernovae. {\it Nature} {\bf 2}, 562 (2006)


\bibitem{rt.temp} T.T. Clark.  A numerical study of the statistics of a two dimensional Rayleigh-Taylor mixing layer. {\it Phys. Fluids} {\bf 15}, 2413 (2003)

\bibitem{Grad49} H. Grad.  On the kinetic theory of rarefied gases. {\it Pure Appl. Math.} {\bf 2}, 325 (1949)



\bibitem{lele1}F. Belletti, L. Biferale, F. Mantovani, S. F. Schifano,
F. Toschi \& R. Tripiccione. Multiphase Lattice Boltzmann on the Cell
Broadband Engine. {\it Il Nuovo Cimento C} {\bf 32} 53 (2009). 

\bibitem{lele}  L. Biferale, F. Mantovani, M. Sbragaglia,
  A. Scagliarini, S. F. Schifano,
F. Toschi, R. Tripiccione. High resolution study of compressible \RT\
turbulence. in preparation (2010). 










\end{thebibliography}
\end{document}